\documentclass[a4paper, amsfonts, amssymb, amsmath, showkeys, floatfix, nofootinbib, twoside, superscriptaddress, aps, prl,reprint]{revtex4-2}
\usepackage[american]{babel}

\usepackage[dvipsnames]{xcolor}
\usepackage{xfrac}
\usepackage{siunitx}
\usepackage{gensymb}
\usepackage{array}
\usepackage{longtable}

\usepackage{physics}

\usepackage{mathtools}
\usepackage[version=4]{mhchem}
\usepackage{booktabs}
\usepackage{makecell}

\DeclareSIUnit{\gauss}{G}
\usepackage[pdftex, pdftitle={Article}, pdfauthor={Author}]{hyperref} 
\usepackage[normalem]{ulem}

\begin{document}

\title{Observation of hysteresis in an isolated quantum system of disordered Heisenberg spins}

\author{Moritz Hornung}
\thanks{these authors contributed equally to this work }
\affiliation{Physikalisches Institut, Universit\"at Heidelberg, Im Neuenheimer Feld 226, 69120 Heidelberg, Germany}

\author{Eduard J. Braun}
\thanks{these authors contributed equally to this work }
\affiliation{Physikalisches Institut, Universit\"at Heidelberg, Im Neuenheimer Feld 226, 69120 Heidelberg, Germany}
\author{Sebastian Geier}
\affiliation{Physikalisches Institut, Universit\"at Heidelberg, Im Neuenheimer Feld 226, 69120 Heidelberg, Germany}
\author{Titus Franz}
\affiliation{Physikalisches Institut, Universit\"at Heidelberg, Im Neuenheimer Feld 226, 69120 Heidelberg, Germany}
\affiliation{Munich Center for Quantum Science and Technology (MCQST), Schellingstr. 4, D-80799 München, Germany}
\affiliation{Max-Planck-Institute for Quantum Optics, Hans-Kopfermann-Str.1, Garching D-85748, Germany}
\author{Gerhard~Z\"urn}
\affiliation{Physikalisches Institut, Universit\"at Heidelberg, Im Neuenheimer Feld 226, 69120 Heidelberg, Germany}
\author{Matthias~Weidem\"uller}
\email{weidemueller@uni-heidelberg.de}
\affiliation{Physikalisches Institut, Universit\"at Heidelberg, Im Neuenheimer Feld 226, 69120 Heidelberg, Germany}

\date{\today}

\begin{abstract}

We find energy-dependent hysteresis in an
isolated Heisenberg quantum spin system, similar to thermomagnetic hysteresis in canonical spin glasses in contact with a thermal reservoir. 
Analogous to zero-field cooling and field cooling in conventional magnetic materials, an annealing protocol is devised to control the energy in an isolated system.
Depending on the strength of disorder, the susceptibilities at zero field bifurcate at a specific energy, which signals the presence of different magnetic regimes.
This behavior is apparent both in a numerical simulation by exact diagonalization of the Heisenberg Hamiltonian with twelve particles, as well as in an experiment with thousands of Rydberg atoms representing dipolar interacting quantum spins. The annealing protocols open a new path to explore the energy-dependent phase structure of spin systems at low energies.
Our observation of a nonthermal metastable regime might indicate the existence of a phase transition to a novel state of isolated quantum spins.

\end{abstract}

\maketitle

Frustrated magnetism is defined by the absence of a unique spin configuration that would minimize the competing interaction energies of all bonds \cite{fisher_theory_1988}. It leads to the formation of novel exotic phases such as spin ices \cite{castelnovo_spin_2012, snyder_how_2001, bramwell_history_2020, castelnovo_magnetic_2008}, spin liquids \cite{kivelson_50_2023, zhou_quantum_2017,cepas_heterogeneous_2012,semeghini_probing_2021} and super spin glasses \cite{de_toro_ideal_2014,mathieu_phase_2013,nadeem_effect_2011}. The first model system where magnetic frustration was extensively studied is the spin glass \cite{binder_spin_1986,vannimenus_theory_1977, charbonneau_spin_2023, parisi_nobel_2023,mydosh_spin_2014,mydosh_spin_2015}, which developed into a 
paradigmatic model to study many-body problems ranging from non-ergodicity and nonthermal behavior \cite{baldwin_clustering_2017,laumann_many-body_2014} to quantum optimization problems \cite{king_quantum_2023,bernaschi_quantum_2024, leschke_existence_2021}. In isolated quantum systems, the existence of a spin glass phase is well established for Ising-like models \cite{rieger_quantum_1997,harris_phase_2018,rademaker_slow_2020,hosseinabadi_quantum--classical_2024,marsh_entanglement_2024}. For the specific case of dipolar interactions, the nature of the spin glass phase is still under debate even in classical spin systems \cite{reich_dipolar_1990,ghosh_coherent_2002,schechter_quantum_2007,biltmo_unreachable_2012,zhang_chiral-glass_2011,sen_topological_2015,bose_dipolar_2019}.

Common protocols to study spin glass behavior in solid state physics are magnetic response measurements upon zero-field cooling (ZFC) and field cooling (FC) \cite{das_observation_2019,mugiraneza_tutorial_2022,joy_comparison_2000,peixoto_analysis_2017,parisi_nobel_2023}.
In both protocols, the sample is initially prepared in its high-temperature paramagnetic phase. In the zero-field cooled (ZFC) protocol, the sample is first cooled to the target temperature in the absence of an external magnetic field. A small probe field is then applied, and the resulting magnetization along the field direction is measured to determine the magnetic susceptibility. In the field cooled (FC) protocol, the same probe field is applied throughout the cooling process, and the magnetization is measured upon reaching the target temperature.
Accordingly, the ZFC protocol probes the linear-response susceptibility of the low-temperature equilibrium state in zero field. In contrast, the FC protocol measures the magnetization of thermal equilibrium states established in the presence of a finite external field, thus probing the field-dependent equilibrium response \cite{parisi_nobel_2023}.

If the system remains in thermodynamic equilibrium throughout the cooling process, the susceptibilities measured in the ZFC and FC protocols coincide. A bifurcation of the two susceptibilities is a hallmark of thermomagnetic hysteresis, reflecting the dependence of the system on its thermal and magnetic history.
The temperature at which the bifurcation appears defines the onset of irreversibility and is commonly used as a proxy for the spin glass transition temperature \cite{kenning_irreversibility_1991,vincent_ageing_2007, pradhan_investigation_2024}.

\begin{figure}[!t]
    \centering
    \includegraphics[]{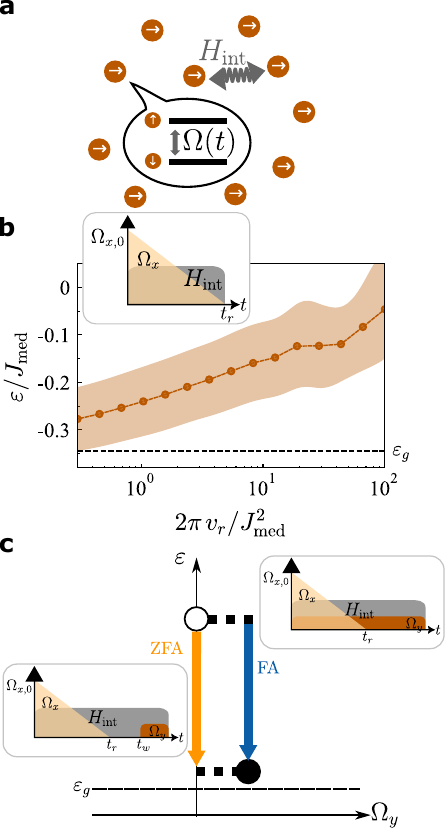}
    \caption{(a) Interacting spin system governed by the Hamiltonian \(H_\mathrm{int}\). Each spin represents a driven two-level system with external field \(\Omega(t)\).
    (b) Annealing scheme for the dipolar spin system. Inset: The annealing field \(\Omega_x\) (light orange) is linearly ramped from \(\Omega_{x,0}\) to zero over a time \(t_r\) in the presence of the interaction Hamiltonian \(H_\mathrm{int}\) (gray). Main: Mean interaction energy per particle \(\varepsilon\) after the ramp as a function of the ramp speed \(v_r=\Omega_{x,0}/t_r\). The shaded region denotes the standard deviation, and the dashed line the mean ground-state energy \(\varepsilon_g\).
    (c) Comparison of the ZFA (left inset) and FA (right inset) protocols. In ZFA, the system equilibrates for a waiting time \(t_w\) before the probe field \(\Omega_y\) (brown) is applied; in FA, \(\Omega_y\) is present throughout the sequence. Main: Accessible phase space spanned by \(\varepsilon\) and \(\Omega_y\). Starting from the fully polarized state (black circle on abscissa),the low energy state at finite \(\Omega_y\) (filled black circle) can be prepared at identical \(\varepsilon\) and \(\Omega_y\). The ZFA protocol anneals the energy at zero field (orange arrow), while the FA protocol anneals the energy at finite \(\Omega_y\) (blue arrow). Quenches of the probe field are indicated by black dotted lines.
    }
    \label{fig:Fig.1}
\end{figure}

In this Letter, we adapt these protocols that typically require coupling to a thermal reservoir to the regime of isolated quantum systems by introducing an annealing procedure that controls the mean interaction energy. We implement both zero-field annealing (ZFA) and field annealing (FA) in numerical simulations and in a Rydberg atom quantum simulator for a disordered dipolar Heisenberg-XY model in three dimensions. Our results reveal an energy-dependent hysteresis behavior, reminiscent of thermomagnetic hysteresis observed in canonical spin glasses \cite{wolf_relaxation-caused_1993,mamiya_comparison_2007,mahmud_field_2017}.

Our newly devised experimental protocols rely on two time-dependent external fields: an annealing field \(\Omega_x\) used to tune the energy in the system and a probe field \(\Omega_y\) used to extract the magnetization response.
To be specific, we concentrate on the Hamiltonian \(\left(\hbar=1\right)\) governing a disordered dipolar Heisenberg-XY model of \(N\) spins
\begin{equation}
    \label{eq:hamiltonian}
    \begin{split}
    H &= H_\mathrm{int} + H_\mathrm{para}\\
    &=\underbrace{\sum_{i<j} J_{ij} \left(S_x^{(i)}S_x^{(j)}+S_y^{(i)}S_y^{(j)}\right)}_{H_\mathrm{int}} \underbrace{-\sum_{i=1}^N\left(\Omega_xS^{(i)}_x+\Omega_yS^{(i)}_y\right)}_{H_\mathrm{para}},
    \end{split}
\end{equation}
as schematically depicted in Fig. \ref{fig:Fig.1}(a). \(S_\mathrm{\alpha}^{(i)}\) represents the component \(\alpha\) of the spin operator acting on spin \(i\) and \(J_{ij}\) the coupling strength between spins \(i\) and \(j\). The couplings follow a dipolar interaction law \(J_{ij}=\frac{C_3}{r_{ij}^3}(1-3\cos^2(\theta_{ij}))\), where \(C_3\) is the dispersion coefficient, \(r_{ij}\) the distance between spins \(i\) and \(j\), and \(\theta_{ij}\) the angle between the interparticle axis and the \(z\)-axis. 
As the timescale of the dynamics in the system is mainly determined by the strength of nearest neighbor couplings, disregarding their sign, we introduce the median of the distribution \( J_\mathrm{med}\coloneqq \underset{i}{\mathrm{median}}\: \underset{j}{\max}\left|J_{ij}\right|\) as a reference unit of coupling strength \cite{franz_observation_2024}.

To tune the mean interaction energy of a time-dependent state \(\ket{\Psi(t)}\)
\begin{equation}
    \varepsilon = \frac{1}{N}\expval{H_\mathrm{int}}{\Psi(t)}
\end{equation}in such a long-range interacting spin system, we apply the annealing scheme shown in the inset of Fig. \ref{fig:Fig.1}(b). The spin system is prepared at \(t=0\) in the fully polarized spin eigenstate in \(x\)-direction \(\ket{\Psi_x}=\ket{\rightarrow}^{\bigotimes N} \) with \(S_x\ket{\rightarrow}=\frac{1}{2}\ket{\rightarrow}\), such that \(\varepsilon(0)\approx 0\). To tune \(\varepsilon\), a time-dependent annealing field is applied along the spin direction, following \(\Omega_x(t) =  \Omega_{x,0} - v_r t ,\quad 0\leq t\leq t_r\), 
where \(\Omega_{x,0}\) denotes the initial amplitude of the annealing field and \(v_r = \Omega_{x,0} /{t_r}\) the ramp speed with the ramp time \(t_r\).

\begin{figure*}[!ht]
    \centering
    \includegraphics[width=\linewidth]{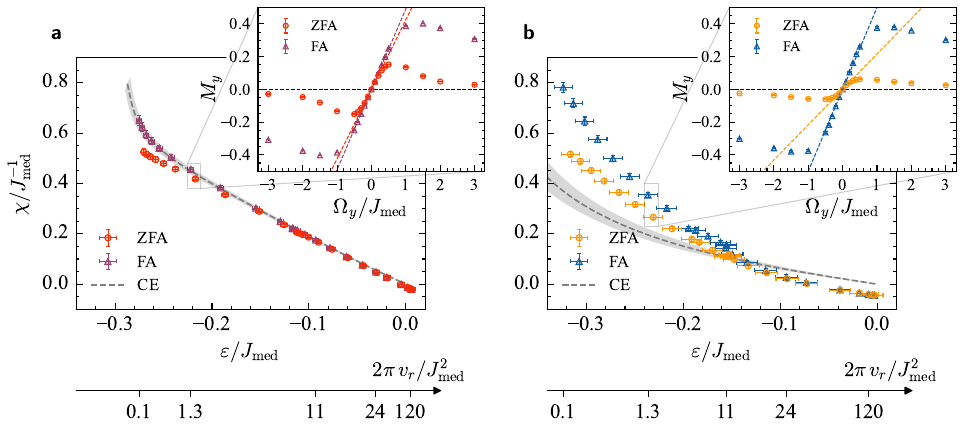}
    \caption{Exact diagonalization of Eq. \ref{eq:hamiltonian} with \(N=12\) spins,  \(\Omega_{x,0}=2.4 J_\mathrm{med}\) and \( \Omega_y=0.1 J_\mathrm{med}\) in the case of (a) weakly disordered and (b) strongly disordered spin configurations. Disorder is quantified by the standard deviation of the nearest-neighbor coupling distribution, which is \(\sigma_{\mathrm{J}}/J_{\mathrm{med}}=0.33\) for the weakly disordered case (\(J_\mathrm{med}=4.85\)), and \(\sigma_{\mathrm{J}}/J_{\mathrm{med}}=1.66\) for the strongly disordered case (\(J_\mathrm{med}=0.72\)). In each case, linear magnetic susceptibility \(\chi_{\mathrm{mag}}\) is plotted for both the ZFA and FA protocol as a function of the energy per particle $\varepsilon$ in the final state after the ramp. The corresponding ramp speed is given on the lower, nonlinear axis. In the weak disorder case, only a small difference between the protocols is apparent and the susceptibility can be well described by a canonical ensemble prediction (grey shaded area). In the strong disorder case, the thermal prediction fails and there is a clear bifurcation for low energies. The errorbars are the standard error of the mean resulting from a disorder average of 100 (250) realizations for weak (strong) disorder. The insets display magnetization \(M_y\) versus probe field \(\Omega_y\) at fixed ramp speed \(2\pi v_r\approx J_{\mathrm{med}}^2\) for both ZFA and FA protocols. $\chi_\mathrm{mag}$ is derived from the slope of this magnetization curve  around \(\Omega_y=0\).
    }
    \label{fig:Fig.2}
\end{figure*}

If \(\Omega_{x,0}\) is large compared to all \(J_{ij}\), spin-spin interactions can be neglected at \(t=0\) and the Hamiltonian Eq. \ref{eq:hamiltonian} reduces to the paramagnetic Hamiltonian \(H_\mathrm{para}\).
In this paramagnetic regime, the ground state is the fully polarized state \(\ket{\Psi_x}\).
The field ramp can be interpreted as an annealing scheme from \(H_\mathrm{para}\) to \(H_\mathrm{int}\) \cite{hauke_perspectives_2020}. Variation of the ramp speed \(v_r\) allows one to initialize states at different \(\varepsilon(t_r)\), as depicted in Fig. \ref{fig:Fig.1}(b).
Fast (diabatic) ramps leave the system in the initial polarized state \(\ket{\Psi_x}\), corresponding to a high-energy state with \(\varepsilon \sim 0\). In contrast, slow (adiabatic) ramps allow the system to follow the instantaneous eigenstate, and thus prepare a state with an energy close to the energy of the ground state \(\varepsilon_g\).

The states at tunable \(\varepsilon\) can be prepared and probed using two protocols, as illustrated in Fig. \ref{fig:Fig.1}(c). In the zero-field-annealed (ZFA) protocol, the system is allowed to equilibrate at \(\Omega_x=0\) for a waiting time \(t_w\) after the annealing ramp, during which the magnetization relaxes to a steady state (see Fig. 7 of the Supplemental Material \cite{supplement}). A small probe field \(\Omega_y\) is then applied, and the resulting steady-state magnetization \(M_y\) is measured along the \(y\)-direction. The probe field is chosen in a direction orthogonal to the annealing field, such that the measured magnetization is generated solely by \(\Omega_y\). The ZFA susceptibility therefore probes the linear response of the post-annealing state to a weak transverse field.

In the field-annealed (FA) protocol, the probe field \(\Omega_y\) is applied throughout the annealing ramp and remains on while the system equilibrates. The FA protocol thus probes the magnetization of differently prepared low-energy states in the presence of a finite probe field. 

For both protocols, the magnetization is linearized at \(\Omega_y=0\), yielding the zero-field susceptibility
\begin{equation}
    \chi \coloneqq \left.\frac{\partial M_y}{\partial \Omega_y}\right|_{\Omega_y=0}.
\end{equation}
For an ETH thermal system, the susceptibilities in both the ZFA and FA protocols are expected to coincide, since equilibrium observables depend only on the mean interaction energy and the applied magnetic field. Therefore, a bifurcation between the ZFA and FA susceptibilities constitutes a signature of nonthermal behavior.

We simulate the full quantum dynamics of the proposed protocols numerically on a system of twelve spins governed by \(H_\mathrm{int}\) for two different disorder configurations. The disorder is introduced by sampling the spin positions uniformly within a unit sphere while enforcing a minimal interspin distance \( r_\mathrm{bl} \).
This positional disorder generates a distribution of coupling strengths, which we quantify using the relative standard deviation \(\sigma_{\mathrm{J}} / J_{\mathrm{med}}\), with \(\sigma_{\mathrm{J}} \coloneqq \underset{i}{\mathrm{std}}\: \underset{j}{\max}\left|J_{ij}\right|\).
Choosing \( r_\mathrm{bl} = \num{0.39} \) and \( r_\mathrm{bl} = \num{0.12} \) yields weakly ($\sigma_\mathrm{J}/J_\mathrm{med}=0.33$) and strongly ($\sigma_\mathrm{J}/J_\mathrm{med}=1.66$) disordered configurations, respectively. The coupling distributions are further detailed in the Supplemental Material \cite{supplement}).

The insets of Fig. \ref{fig:Fig.2} present the ZFA and FA magnetization for a fixed ramp speed as a function of the probe field, revealing a linear response regime at small probe field amplitudes.
The susceptibilities for the ZFA and FA protocols are extracted from the slope of the magnetization around zero field and shown in Fig. \ref{fig:Fig.2} as a function of the mean interaction energy per particle and the corresponding ramp speed.
We observe that the magnetization vanishes for mean interaction energies close to \( \varepsilon \approx 0 \), and increases toward lower energies. While the ZFA and FA susceptibilities coincide for almost all energies in case of the weak disorder configuration, we observe a clear bifurcation in case of the strong disorder configuration at \( \varepsilon / J_\mathrm{med} \approx -0.15 \), clearly indicating nonthermal behavior. 

This motivates a comparison with predictions from the canonical ensemble (CE). In the weakly disordered configuration, shown in Fig. \ref{fig:Fig.2}a, the ZFA and FA susceptibilities coincide with the CE prediction except at energies close to the ground state, indicating thermal behavior. The agreement of the FA susceptibility with the CE prediction over the entire energy range further supports ETH thermalization.
In contrast, in the strongly disordered configuration shown in Fig. \ref{fig:Fig.2}b, neither the ZFA nor the FA susceptibility is described by the CE, indicating that both protocols prepare metastable non-equilibrium states.

To interpret these findings, we begin by considering a paramagnetic system. As shown in the Supplemental Material \cite{supplement}, an isolated paramagnet exhibits no response. Motivated by previous work showing that the dynamics of strongly positionally disordered spin systems are often well described by independent spin pairs \cite{signoles_glassy_2021,braemer_pair_2022,christakis_probing_2023,franz_observation_2024}, we investigate an ensemble of non-interacting spin pairs, each governed by an XY Hamiltonian with a different coupling strength \(J\). As detailed in the Supplemental Material \cite{supplement}, this simple picture captures the essential features of our observations.

For slow ramps, the annealing protocol prepares each pair in its ground state in the presence of an infinitesimal field perpendicular (parallel) to the probe field for the ZFA (FA) protocol. Since the ground state exhibits a larger magnetic response than the excited state, the susceptibility increases for slower ramps, which prepare lower-energy states. Moreover, for a distribution of coupling strengths, slower ramps drive a larger fraction of spin pairs into their ground state, further enhancing the susceptibility. For positive couplings (\(J>0\)), the response along the probe-field direction differs between the annealed ground states prepared by the two protocols, which leads to the observed bifurcation between the ZFA and FA susceptibilities. Both slower ramps and stronger disorder increase the fraction of pairs with \(J>0\) that reach their ground state during annealing, explaining why a pronounced bifurcation emerges only in the regime of low energy and strong disorder. Since the spin pairs are non-interacting, the dynamics decomposes into independent spin pairs and is therefore not expected to thermalize \cite{franz_absence_2022}. This naturally explains why the canonical ensemble fails to describe both the ZFA and FA protocols in the strong-disorder limit.

\begin{figure*}[!t]
    \centering
    \includegraphics[width=\linewidth]{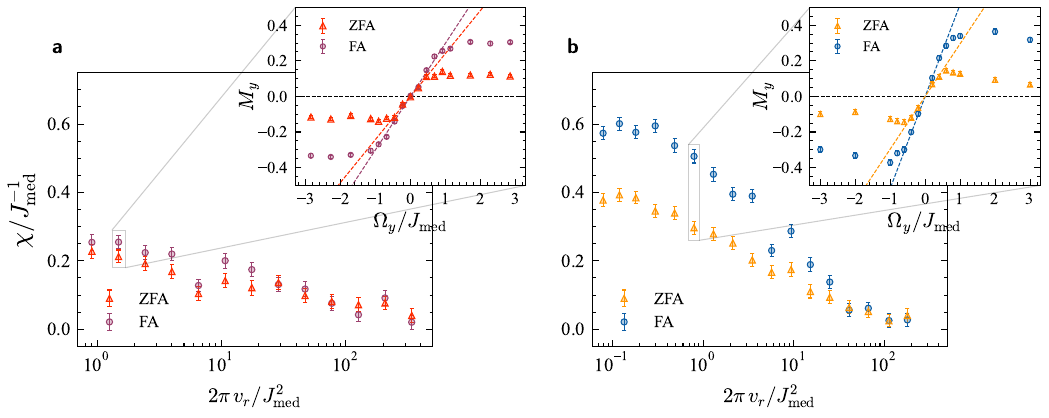}
    \caption{(a) Linear susceptibility \(\chi_\mathrm{mag}\) in units of \(J_\mathrm{med}\) as a function of the ramp speed \(v_r\) rescaled by \(J_{\mathrm{med}}\) = \(2\pi \times \SI{1.01}{\mega\hertz}\) for the ZFA (red) and FA (violet) magnetization in the weakly disordered configuration. Inset shows response measurements of \(y\)-magnetization as a function of the dimensionless probe field \(\Omega_y/J_\mathrm{med}\). Magnetization is measured \SI{3}{\micro \second} after the end of the annealing ramp, corresponding to 5 interaction cycles of the median interaction. This ensures both protocols have the same absolute duration independent of the waiting time in the ZFA protocol.
    The dashed lines indicate a linear fit through the origin to the five data points centered around zero, used to extract the susceptibility and its error bar. (b) Linear susceptibility \(\chi_\mathrm{mag}\) in units of \(J_\mathrm{med}\) as a function of the ramp speed \(v_r\) rescaled by \(J_{\mathrm{med}}\) = \(2\pi \times \SI{1.63}{\mega\hertz}\) for the ZFA (yellow) and FA (blue) magnetization in the strongly disordered configuration. Inset shows response measurements of \(y\)-magnetization as a function of the dimensionless probe field \(\Omega_y/J_\mathrm{med}\). The dashed lines indicate a linear fit through the origin to the four data points centered around zero, used to extract the susceptibility and its error bar.
    }
    \label{fig:Fig.3}
\end{figure*}

In order to investigate whether these findings also persist for a larger system size consisting of thousands of spins, we experimentally implement the ZFA and FA protocols on a Rydberg atom quantum simulation platform that realizes the Heisenberg XY-model \cite{geier_floquet_2021,franz_absence_2022,franz_observation_2024}.
Here, \ce{^{87}Rb} atoms are loaded into a crossed-beam optical dipole trap at \SI{11}{\micro\kelvin} in the ground state \(\ket{5S_{1/2},F=2,m_F=2}\). 
After release, the atomic cloud expands freely, forming an approximately Gaussian distribution of \num{10300} (\num{9300}) atoms with an axial width of \SI{140}{\micro\meter} (\SI{137}{\micro\meter}) and a radial width of \SI{49}{\micro\meter} (\SI{97}{\micro\meter}) in the weakly (strongly) disordered configuration. 
The emergence of these disorder configurations depends on both the cloud parameters and the Rydberg excitation scheme, and will be discussed below.

From this cloud, approximately \SI{20}{\percent} of the atoms are excited via a two-photon transition to the Rydberg state \(\ket{61S_{1/2},m_J=1/2}=\vcentcolon \ket{\downarrow}\) to prepare a frozen Rydberg gas.
By application of a resonant microwave drive from \(\ket{\downarrow}\) to a \(\ket{61P}=\vcentcolon \ket{\uparrow}\) state, an effective 2-level system can be realized, which naturally implements the Heisenberg XY-Hamiltonian \cite{franz_observation_2024}. The phase and amplitude of the microwave drive can be changed with an arbitrary waveform generator to implement the annealing and probe fields in the Hamiltonian Eq. \ref{eq:hamiltonian}.

The positional disorder of the Rydberg atoms, as described in previous work \cite{signoles_glassy_2021,franz_observation_2024}, originates from the Gaussian distribution of the ground-state atoms and is modified by the Rydberg blockade \cite{lukin_dipole_2001,gaetan_observation_2009,urban_observation_2009} with \( r_\mathrm{bl} \approx \SI{8}{\micro\meter} \). The blockade arises from the van-der-Waals interaction during excitation, characterized by a dispersion coefficient \( C_6/(2\pi) = \SI{188}{\giga\hertz \, \micro\meter^6} \), which becomes anisotropic in the presence of a strong magnetic field.
This positional disorder translates into a distribution of couplings \( J_{ij} \) in the XY Hamiltonian in Eq. \ref{eq:hamiltonian}. While the experimental coupling distribution differs quantitatively from the 12-particle numerical simulation discussed above due to the Gaussian atomic cloud, it reproduces the same qualitative features (see Supplemental Material \cite{supplement}).
In the weakly (strongly) disordered configuration, the median coupling \( J_\mathrm{med}/(2\pi) \) is \SI{1.01}{\mega\hertz} (\SI{1.63}{\mega\hertz}), with disorder quantified by \( \sigma_J / J_\mathrm{med} = 1.02 \) (\(2.24\)). Further details on the experimental implementation are provided in the Supplemental Material \cite{supplement}.

Figure \ref{fig:Fig.3} shows the measured ZFA and FA susceptibilities for different ramp speeds in both disorder configurations. We observe qualitatively the same behavior as in the few-particle simulations, corroborating the independent-pair description, whose disorder-averaged response is insensitive to the system size. At \(2\pi v_r \approx 10J_\mathrm{med}^2\), the measured susceptibilities exhibit a local minimum as a function of ramp speed that is absent in Fig. \ref{fig:Fig.2}. However, the same feature is recovered in the few-particle simulations when the susceptibilities are plotted as a function of ramp speed. It originates from the nonlinear dependence of the energy on the ramp speed (see Supplemental Material \cite{supplement}).

The energy-dependent hysteresis shares several features with canonical spin glasses, while also exhibiting important differences. As in the separation between zero-field cooled (ZFC) and field cooled (FC) responses, the difference between the FA and ZFA susceptibilities grows at lower energies and becomes more pronounced with increasing disorder. The existence of two distinct states with different susceptibilities, selected by the preparation protocol, is likewise reminiscent of the thermomagnetic hysteresis between ZFC and FC states.
However, the observed behavior differs from canonical spin glasses in several aspects. In particular, both FA and ZFA susceptibilities continue to increase monotonously after the bifurcation toward lower energies, whereas in canonical spin glasses the FC susceptibility saturates and the ZFC susceptibility decreases. Moreover, hysteresis in the strongly disorder configuration appears already at relatively high energies and for fast ramps.
In fact, we observe both a (meta)stable magnetization and strong hysteresis after a few \(J_\mathrm{med}^{-1}\), suggesting that the relevant dynamics and timescales are fundamentally different from thermally activated relaxation processes in canonical spin glasses.
Similar to dipolar Ising Spin glasses \cite{brooke_quantum_1999,ghosh_coherent_2002}, quantum tunneling may therefore play a role in the formation of these non-equilibrium states.

The observed hysteresis demonstrates that the FA and ZFA protocols prepare states with a memory of the annealing history, indicating the presence of non-equilibrium states. Since at least one of the states prepared by the two protocols is out of thermal equilibrium, these protocols provide access to the emergence and characterization of nonthermal states in isolated quantum systems. This raises the question of how such memory effects relate to thermalization. Similar questions arise in spin glass systems, where the nature of field cooled and zero-field cooled states and their relation to equilibrium phases remain actively debated \cite{parisi_nobel_2023}. Recently, a nonthermal FC state has been identified in superspin glasses \cite{pal_field-cooled_2020,pal_non-equilibrium_2023}, which are also governed by dipolar interactions.

Because the FA and ZFA protocols devised in this letter are not limited to systems with dipolar interactions, they provide a general framework for probing thermalization and non-equilibrium states in a wide range of quantum systems. In particular, their application to experimentally realized quantum spin glass systems with all-to-all interactions \cite{king_quantum_2023,kroeze_directly_2025} may shed new light on the long-standing question of the nature of the ZFC and FC states and their relation to the ZFA and FA states. Beyond these fundamental questions, the protocols also provide a practical tool for benchmarking adiabaticity and performance in annealing-based quantum computing platforms \cite{grant_benchmarking_2021,subires_benchmarking_2022,king_quantum_2023,kim_quantum_2025}.

\begin{acknowledgments}
We thank A. Braemer, M. Gärttner, S. de Léséleuc and M. Müller for fruitful discussions. E. J. Braun acknowledges support by the IMPRS for Quantum Dynamics in Physics, Chemistry and Biology. This work is part of and supported by the Deutsche Forschungsgemeinschaft (DFG, German Research Foundation) under Germany’s Excellence Strategy EXC2181/1-390900948 (the Heidelberg STRUCTURES Excellence Cluster), within the Collaborative Research Centre “SFB 1225 (ISOQUANT)”, the DFG Priority Program “GiRyd 1929,”
the Horizon Europe programme HORIZON-CL4-2022-QUANTUM-02-SGA via
the project 101113690 (PASQuanS2.1), and the Heidelberg Center for Quantum Dynamics. 

\end{acknowledgments}
\appendix

\bibliographystyle{apsrev4-2}
\bibliography{bib}

\end{document}


\section{Exact diagonalization of the few particle system}
\subsection{Finite size analysis}
Motivated by the Rydberg blockade, the $N$ spins are modeled as hard spheres of unit volume in three dimensional space with positions drawn from a uniform distribution over a sphere. In order to create different disorder regimes, the volume $V_\mathrm{tot}$ underlying the distribution is adapted such that a specified packing fraction $N/V_\mathrm{tot}$ is reached. The positions are then used to construct the Heisenberg XY-Hamiltonian presented in the main text. The numerical value of the \(C_3\) coefficient was set to 1 and time is rescaled by the calculated value of \(J_\mathrm{med}\). The initial state for the simulations is always the fully polarized state in \(x\)-direction with the initial field amplitude being set to \(\Omega_{x}=2.4J_\mathrm{med}\). We ran the simulation of the FA and ZFA protocols as described in the main text for $N=4,6,8,10$ and 12 particles at a probe field strength of $\Omega_{y}=0.1J_\mathrm{med}$, the susceptibility is obtained by dividing the steady state magnetization by the applied field. Finally, all simulations were run repeatedly over 100 different disorder realizations and the final plots are disorder averaged and plotted with standard error of the mean, which is in some cases smaller than the indicator and therefore not visible.

\begin{figure}[t]
    \centering
    \includegraphics[width=\linewidth]{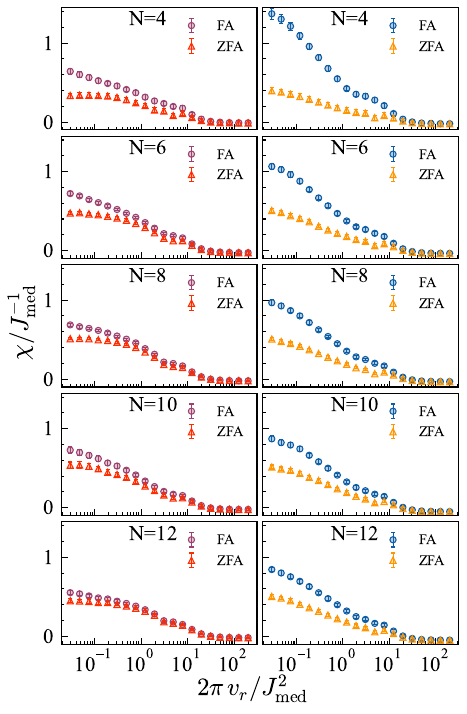}
    \caption{Numerical simulation of FA and ZFA depending on the number of particles in the weakly (left) and strongly (right) disordered regime. }
    \label{fig:FiniteSizeAnalysis}
\end{figure}
The results are shown in Fig.~\ref{fig:FiniteSizeAnalysis}. The left panel shows the weak-disorder regime, and the right panel shows the strong-disorder regime. As expected, the curves differ substantially between $N = 4$ and $N = 12$, reflecting finite-size effects. In the weakly disordered regime, the difference between the curves appears to decrease as $N$ increases. In contrast, a substantial gap remains in the strongly disordered regime, even at $N = 12$. Despite the limited system sizes considered, the numerical results qualitatively capture the dynamics of the experimental system.

\subsection{Energy as a function of ramp time}\label{EnergyAsFunctionOfRampTime}
\begin{figure}[!ht]
    \centering
    \includegraphics[width=\linewidth]{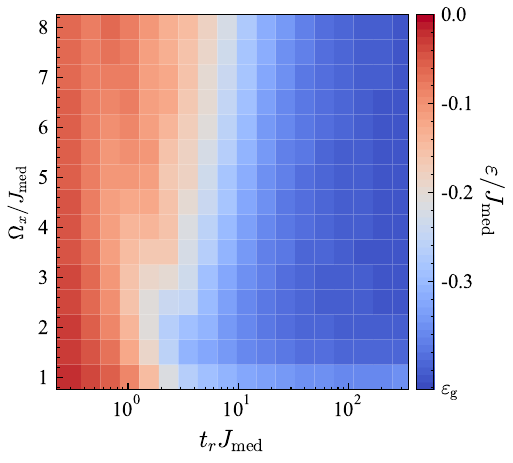}

    \vspace{.5cm}
    \caption{Mean energy per particle after the ZFA ramp ($\Omega_y=0.1J_\mathrm{med}$) as a function of the initial amplitude of the external field $\Omega_x$ and ramp time $t_r$. As long as the initial field amplitude is strong enough ($\Omega_x \gtrsim 2.5J_\mathrm{med}$), the energy is mostly determined through the ramp time.}
    \label{fig:EnergyAfterRamp}
\end{figure}

\begin{figure}[!ht]
    \centering
    \includegraphics[width=.85\linewidth]{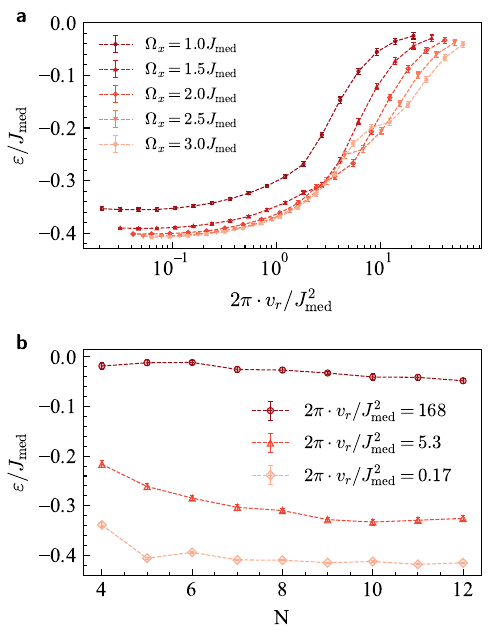}
    \caption{(a) ZFA mean energy per particle as a function of ramp speed for different initial field amplitudes. (b) Behavior of the ZFA mean energy per particle for different system sizes.}
    \label{fig:EnergyFiniteSizeAnalysis}
\end{figure}

Throughout this work, we use a quantum annealing approach to initialize our system at different internal energies. Accordingly, the final energy of the system is determined by the rate at which the Hamiltonian changes during this process. By varying the annealing speed, we are hence able to initialize the system over a large range of different mean energies per particle, which is explicitly shown in the following using an exact diagonalization numerical simulation.

For a broad range of parameters, we simulate the full quantum dynamics of the annealing ramp and calculate the energy expectation value of the state at the end of the ramp with respect to the Hamiltonian including the probe field. Figure~\ref{fig:EnergyAfterRamp} shows the resulting mean energy per particle, $\varepsilon = E/N$, averaged over all disorder realizations, color-coded as a function of the initial field $\Omega_x$ and the ramp time $t_r$. Two limiting cases are readily understood. For very short ramp times, the system does not resolve the annealing process, and the Hamiltonian change can be treated as instantaneous. In this quench limit, the state population is distributed evenly across the spectrum, and the mean energy per particle approaches zero. In the opposite limit of very slow ramps, $t_r J_{\mathrm{med}} \sim 100$, the evolution can be considered adiabatic, and the system remains in the instantaneous ground state. Consequently, the mean energy per particle at the end of the ramp approaches the ground-state energy. Provided that the initial state has sufficient overlap with the ground state, this limiting behavior is nearly independent of the initial field strength.

The influence of the initial field becomes more apparent when the mean energy per particle is plotted over the ramp speed with a parametric dependence on the initial field, as shown in Fig. \ref{fig:EnergyFiniteSizeAnalysis}a. If the initial field is of the same strength as the interactions, the overlap with the ground state is too small for efficient annealing and the minimal energy that is reached for slow ramps is still far from ground state energy. However, already an initial amplitude of twice the interaction strength is enough to reach energy levels in the vicinity of the ground state energy $\varepsilon_{g}$. For even larger initial amplitudes, the behavior only changes for fast ramps and the limit of quenching the field is reached in a different way. Importantly, all the curves are monotonous functions with respect to the ramp speed, such that we can interchangeably speak of increasing the energy or the ramp speed.

Finally, Fig. \ref{fig:EnergyFiniteSizeAnalysis}b shows how the influence of the annealing ramp develops as the system size increases. The mean energy per particle after the ramp is plotted as function of the system size for three different annealing speeds that are realistic in regard to possible experimental timescales. In general, the mean energy per particle at fixed ramp speed decreases as the system increases until it eventually appears to saturate for values of $N \geq 10$. With this behavior in mind, we are confident that our approach initializes a wide range of energies even in the many-body case.  
We note that this discussion is purely about the energy expectation value and does not make any statement about the ground state overlap. In this, our approach differs greatly from other annealing applications, where the exact ground state configuration of a system is searched for.

\subsection{Eigenstate picture of the annealing protocol}

\begin{figure}[!ht]
    \centering
    \includegraphics[width=\linewidth]{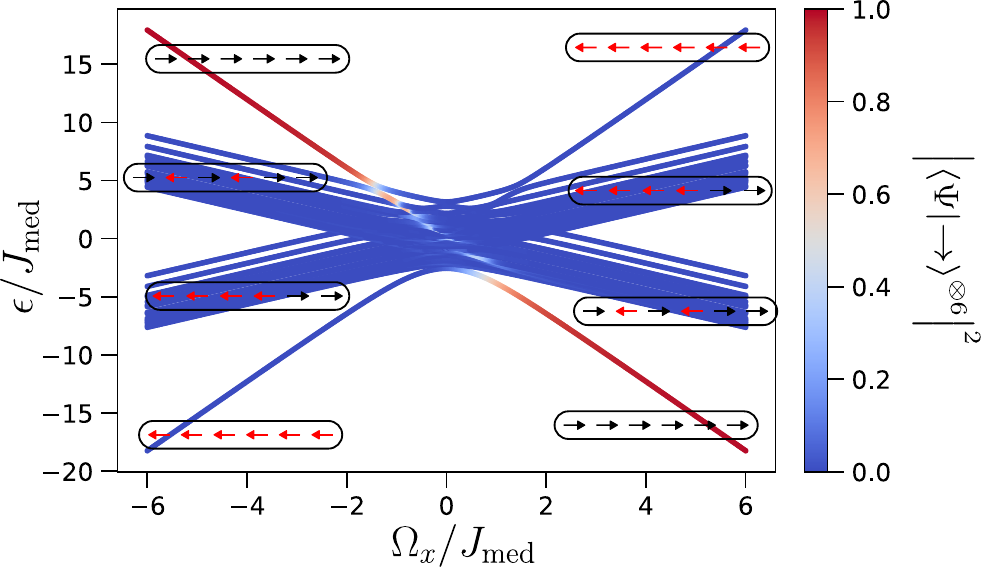}

    \vspace{.5cm}
    \caption{Eigenenergies \(\epsilon\) in the positive parity subspace as a function of the applied magnetic field \(\Omega_x\) for a single disorder realization with 6 particles. All energies are given in units of \(J_\mathrm{med}\). The color indicates the overlap of the eigenvector corresponding to a certain eigenstate with the fully polarized state \(\ket{\rightarrow}^{\otimes 6}\). Arrows indicate asymptotic states for large magnetic fields, where either 0, 2, 4, or all 6 spins are reversed. Aligned spins are depicted in black, while antialigned states are depicted in red.}
    \label{fig:LevelDiagram}
\end{figure}
For a better illustration of the annealing protocol described in the previous section, we explain the change of the energy with respect to the ramp time in depth for the two limiting cases:  an infinitely fast quench and an infinitely slow adiabatic ramp. In order to understand these cases, we will look at the spectrum of a disordered Heisenberg XY model with a transverse field in \(x\)-direction as considered in the main text. For such a Hamiltonian, the parity \(\hat{P}=\prod_{i=1}^N \hat{S}_x^i\) is conserved. As a consequence, states with different parity do not interact with each other. As the annealing protocol starts from the fully polarized state \(\ket{\rightarrow}^{\otimes N}\), which is of even parity for even \(N\), during the annealing protocol only states with even parity can be occupied. Therefore, we will only consider this subspace as our Hilbert space. 

For a single realization with \(N=6\) particles we plot the eigenenergies of the positive parity eigenstates as a function of the applied annealing field \(\Omega_x\) in Fig. \ref{fig:LevelDiagram}. For large fields \(\left|\Omega_x \right| \gg J_{med}\), the change in eigenenergies is linear in the annealing field. In this regime, the spin-spin interaction is a perturbation to the paramagnetic limit, where each spin corresponds to an energy of \(\pm \frac{1}{2} \Omega_x\). As there is always an even number of aligned and antialigned spins due to the parity restrictions, the asymptotic energies are \(6\cdot (-1/2) \Omega_x = -3\Omega_x\), \((4\cdot (-1/2) + 2\cdot 1/2) \Omega_x = -\Omega_x\), \((2\cdot (-1/2) + 4\cdot 1/2) = \Omega_x\) and  \(6\cdot 1/2\Omega_x = 3\Omega_x\). In the figure, the asymptotic spin configuration is depicted next to corresponding eigenenergies.

We observe that for large positive fields, the ground state has almost perfect overlap with the fully spin polarized state. As such, an infinitely slow adiabatic annealing from the fully spin polarized state at large initial field corresponds to asymptotically following the ground state. In this case, when the field is slowly ramped to \(\Omega_x \rightarrow 0\), the ground state of \(H_\mathrm{XY}\) is prepared. In the light of a microcanonical ensemble, the ground state corresponds to a zero temperature state. When \(\Omega_x\) is instead quenched to zero, many different eigenstates are populated, but the states that are most populated are the ones in the middle of the spectrum, close to \(\epsilon=0\). In the light of a microcanonical ensemble, this energy corresponds to the energy of an infinite temperature state. Hence by changing the ramp speed between these to limiting cases, any energy between the ground state energy and zero energy can be achieved.

\subsection{Hysteresis as a function of ramp speed}

\begin{figure}[!t]
    \centering
    \includegraphics[]{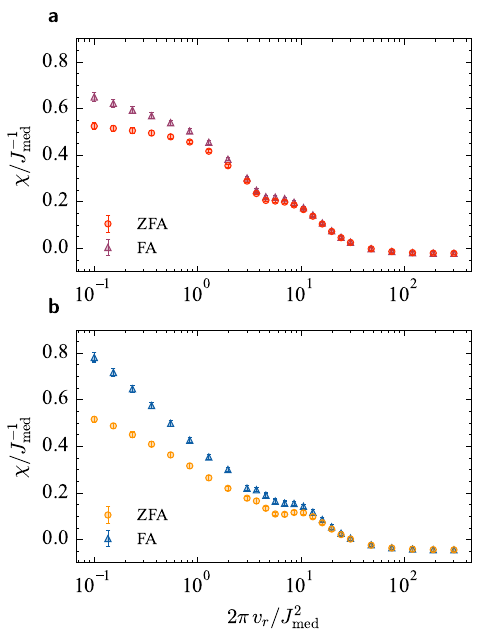}
    \caption{Numerically simulated susceptibilities as a function of the ramp speed for a system of 12 spins. For both the weakly disordered (a) and the strongly disordered case (b), we find a similar peculiarity as observed in the experiment at a ramp speed of $2\pi v_r \approx 10J_\mathrm{med}^2$}
    \label{fig:SuscRampSpeed}
\end{figure}

Up to this point, the numerical susceptibilities have been discussed in terms of the energy of the final state after annealing. Although this choice is natural, since it allows for a direct comparison to the predictions of a thermal ensemble, the energy is not an experimentally accessible quantity. We therefore present here the numerical results as a function of the ramp speed to improve comparability with the experiment.

Fig.~\ref{fig:SuscRampSpeed} shows the numerical data for a system of twelve particles. The susceptibility for the ZFA and FA protocols is plotted as a function of the ramp speed for (a) the weakly disordered and (b) the strongly disordered regime. Similarly to before, the protocols display hysteresis, which is visible for slow ramp speeds. In addition, we find that shortly before the onset of hysteresis at $2\pi v_r \approx 10J_\mathrm{med}^2$, the susceptibility exhibits a saddle point. As highlighted in the main text, this feature is also present in the experimental data at ramp speeds of the same order of magnitude. Its absence when plotting susceptibility over the energy suggests that it is a property of the system dynamics during the ramping procedure rather than an intrinsic ``equilibrium'' property of the system.

\begin{figure}[!t]
    \centering
    \includegraphics[]{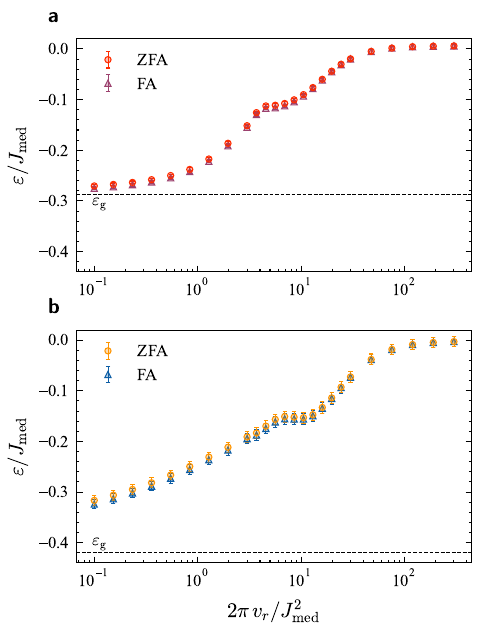}
    \caption{Simulation of the mean energy per particle as a function of the rampspeed in the ZFA and FA protocols in a system of twelve particles. (a) In the weak disorder regime, the ramp initializes states at a variety of energies throughout the full lower half of the spectrum. (b) Same as before but for the strong disorder regime. Again, a wide range of energies is initialized, however, even slower ramps would be needed in order to reach energies in close proximity to the energy of the ground state.}
    \label{fig:EnergyRampSpeed}
\end{figure}

This is confirmed by plotting the energy of the final state as a function of the ramp speed as shown in Fig.~\ref{fig:EnergyRampSpeed}. The overall shape of these is quite similar between the weakly (a) and strongly (b) disordered cases. The energy approaches zero in the limit of an instantaneous quench and decreases with decreasing ramp speed. For both configurations, the curves show a region where the energy is approximately constant with decreasing ramp speed. This occurs for $2\pi v_r \approx 10J_\mathrm{med}^2$, where we previously observed the saddle point in the susceptibility curve. The energy at this speed coincides with the value for which the onset of hysteresis was initially found in the main text. For even slower speeds, the energy continues to decrease toward the ground state energy of the system. In the case of weak disorder, this ground state energy per particle $\varepsilon_{g}\approx0.3J_\mathrm{med}$ is larger than in the case of strong disorder $\varepsilon_{g}\approx 0.4J_\mathrm{med}$. This is due to considerable frustration in the system, which increases the total energy. It is also interesting to note that in the weak disorder case the ground state energy is almost reached, while the system is still quite far from the ground state at strong disorder.

\section{Magnetic response}

\subsection{Interpretation of the response curves through the many-body spectrum}
\begin{figure}[!htpb]
    \centering
    \includegraphics[width=\linewidth]{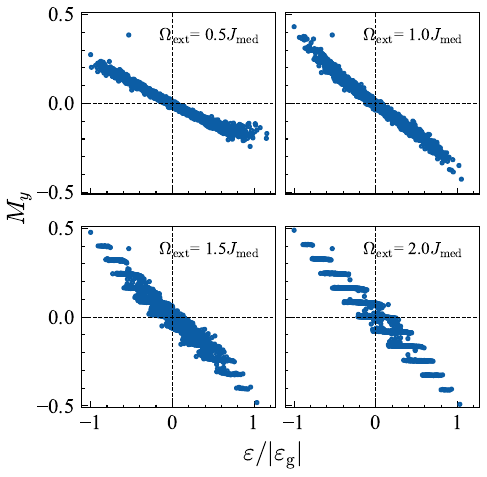}
    \caption{Magnetization spectra for different strengths of the external field. The magnetization of low energy states increases with the external field strength, for fields much larger than $J_\mathrm{med}$, the spectra transforms towards a paramagnetic one where different magnetization sectors separated by a single spin flip become visible.}
    \label{fig:EigenstateMagnetization}
\end{figure}

The goal of this section is to develop an intuitive picture for the shape of the response based on the expectation values of the magnetization for the different eigenstates of the system. To this end, we diagonalize the Hamiltonian at different strengths of the external field and plot the corresponding spectra in Fig. \ref{fig:EigenstateMagnetization}.

To understand the influence of the annealing speed for our system, it is sufficient to consider a single spectrum at fixed field strength. As demonstrated in the previous section, the energy expectation value for slow ramps is very small such that only states in close vicinity to the ground state are populated significantly. From the spectra in Fig. \ref{fig:EigenstateMagnetization}, it becomes apparent that in the presence of a probe field, these states feature non-vanishing magnetization values and therefore also display a magnetic response. On the other hand, as the ramp gets faster and the energy in the system increases, the population is spread out through the spectrum. Higher energy states with negative magnetization values start contributing and cancel out the effect from lower energetic states, eventually yielding zero magnetization.

In order to understand the shape of the response curve at fixed ramp speed, we take a look at how the spectra evolve as the magnetic field changes.
As the amplitude of the probe field increases, it becomes more and more energetically favorable for the spins to align with it. Therefore, low energy states populated through the annealing protocol tend towards larger absolute magnetization values. 
This effect gains in strength with increasing probe field amplitude and leads to the initial increase of magnetization in the response curve. At the same time, in opposition to this mechanism, we find that for both protocols, an increase of the magnetic field strength is always accompanied by a spreading out of the population through the spectrum.

In the FA protocol, this happens due to the initial state, i.e., the fully polarized state in the \(x\)-direction, losing its ground state characteristic once the probe field amplitude in the \(y\)-direction reaches a magnitude similar to the annealing amplitude \(\Omega_{x}\). As a consequence, the initial state is not aligned anymore with the initial field before the ramp and no annealing is possible anymore.

In the ZFA protocol, population is spread out in the process of quenching on strong probe fields large compared to \(J_\mathrm{med}\). Since the magnetization values over the total spectrum are largely symmetric, the magnetization of higher energetic states cancels on average with the magnetization of lower energetic states. 
For large external fields, this effect of spreading out the population becomes predominant, which causes the magnetic response to decline and eventually approach zero as the probe field exceeds a critical value.

\subsection{Linear response of a single particle and a paramagnet at constant energy}

The ZFA and FA magnetizations are obtained by applying a weak probe field to a state at a fixed mean energy per particle. As this state is aligned by construction in the \(x\)-direction at the end of the annealing ramp in the ZFA protocol and at the beginning of the ramp in the FA protocol, the application of the probe field in \(y\)-direction does not introduce new energy to the system. As a consequence, in this section we study the response at constant energy for both a single particle and a paramagnet, which thermalize in a magnetic field in \(x\)-direction. We study the linear response from an equilibrium state with zero magnetization applying the Kubo formula \cite{kubo_statistical-mechanical_1957}. The zero magnetization state is chosen because in the ZFA protocol, the equilibrium state has no net magnetization prior to perturbation. We compare these results to the expected value from classical thermodynamics, assuming an equilibrium state at constant energy.

The Hamiltonian of a single spin interacting with a field in the \(x\)-direction is given as \(H_0=-B\hat{S}_x\). We assume that this single spin reaches thermal equilibrium under this Hamiltonian. We measure the response of the \(y\)-magnetization \(\hat{S}_y\) which is perturbed by the influence of the perturbation Hamiltonian \(H_\mathrm{pert}=-B_y\hat{S}_y\).

In the interaction picture, the time evolution of the operator \(\hat{S}_y\) is given as
\begin{widetext}
\begin{align*}
    \hat{S}_y(t) &= e^{iH_0t}\hat{S}_ye^{-iH_0t} = e^{-iBt\hat{S}_x}\hat{S}_ye^{iBt\hat{S}_x}= \hat{S}_y +\frac{(-iBt)^1}{1!}\comm{\hat{S}_x}{\hat{S}_y}+\frac{(-iBt)^2}{2!}\comm{\hat{S}_x}{\comm{\hat{S}_x}{\hat{S}_y}}+\dots\\
    &= \hat{S}_y+\frac{(Bt)^1}{1!}\hat{S}_z - \frac{(Bt)^2}{2!}\hat{S}_y +\dots= \cos(Bt)\hat{S}_y+\sin(Bt)\hat{S}_z
\end{align*}
\end{widetext}
where we made use of the Hadamard formula in the first line and the spin algebra commutation relation \(\comm{\hat{S}_\alpha}{\hat{S}_\beta}=i\epsilon_{\alpha\beta\gamma}\hat{S}_\gamma\) in the second line. We insert these relations into the Kubo formula:
\begin{widetext}
\begin{align*}
    \expval{\hat{S}_y}(t)&=-i\int_0^t \expval{\comm{\hat{S}_y(t)}{H_\mathrm{pert}(t')}}dt'
    = iB_y\int_0^t \expval{\comm{\cos(Bt)\hat{S}_y+\sin(Bt)\hat{S}_z}{\cos(Bt')\hat{S}_y+\sin(Bt')\hat{S}_z}}dt'\\
    &= -iB_y \int_0^t (\sin(Bt)\cos(Bt')-\sin(Bt')\cos(Bt))i\expval{\hat{S}_x}
    = B_y\expval{\hat{S}_x}\int_0^t \sin(B(t-t'))dt'= B_y \expval{\hat{S}_x} \frac{1-\cos(Bt)}{B},
\end{align*}
\end{widetext}
and thus we obtain for the susceptibility
\begin{equation*}
    \chi(t) = \frac{\expval{\hat{S}_y}(t)}{\partial B_y} = \expval{\hat{S}_x} \frac{1-\cos(Bt)}{B}.
\end{equation*}
We observe that for a state with no equilibrium magnetization, i.e. when \(\expval{\hat{S}_x}\) is zero, the Kubo formula yields no response. In the limit of negligible fields in \(x\)-direction, which is the limit for the ZFA and FA magnetizations at the end of the annealing ramp, we obtain \(\lim_{B\rightarrow0}\frac{1-\cos(Bt)}{B}=0\) and also no response.

The same observations are also true for a paramagnet, as the spin operators have to be only replaced by a sum of the spin operators of each particle. These sums still fulfill the same commutation relations, and hence yield the same result.

This absence of a magnetic response can be understood as well in the light of thermodynamics, where the energy of a paramagnet is given by the relation \(E=-M B\), where \(M\) denotes the magnetization and \(B\) the applied magnetic field. Treating the inner energy as a thermodynamic potential, we get access to the linear susceptibility at constant energy through \(\chi = \left(\frac{\partial M}{\partial B}\right)_{E=\mathrm{const}}\). For a zero magnetic field the energy is zero. So for a constant energy process, the energy will remain zero. As a consequence, we also obtain \(MB = E=0\) for small finite fields, which again yields \(M=0\) for small finite magnetic fields. In a linear regime of external fields around \(B=0\), the resulting magnetization follows \(M=\chi B\). As \(M=0\) also for \(B\neq 0\), this finally yields \(\chi = 0\). In summary, due to the restricted phase space for zero energy, we expect no response.

\subsection{ZFA and FA in the pair picture}
\label{sec:pairpicture}

\subsubsection{Spectrum of a single pair in the presence of a field}

Motivated by previous numerical and experimental studies demonstrating that many dynamical properties of strongly positionally disordered spin systems with power-law interactions can be captured by the dynamics of spin pairs \cite{signoles_glassy_2021,braemer_pair_2022,christakis_probing_2023,franz_observation_2024}, we investigate to what extent the splitting between the ZFA and FA protocols and their energy dependence can be explained by the dynamics of non-interacting pairs. To this end, we first analyze the dynamics of a single pair of spins interacting via a Heisenberg-XY Hamiltonian in the presence of a magnetic field \(\Omega_x\) oriented along the \(x-\)direction, which will play the role of the annealing field in the ZFA protocol
\begin{equation}
    H_0 = J\left(S_x^{(1)}S_x^{(2)}+S_y^{(1)}S_y^{(2)}\right) - \Omega_x \left(S_x^{(1)}+S_x^{(2)}\right).
    \label{eq:pair_hamiltonian_with_field}
\end{equation}
As the XY-Hamiltonian is U(1) symmetric around a rotation in the \(z\)-plane, the following spectral arguments we derive are independent of the direction of the annealing field as long as it is in the \(x-y-\)plane.
For simplicity, we discuss the Hamiltonian in the product basis of the states \(\ket{\pm}_x\) which denote the eigenstates of the spin-1/2 operator \(S_x\) with eigenvalues \(\pm1/2\), respectively.

\begin{figure}
    \centering
    \includegraphics[width=1.\linewidth]{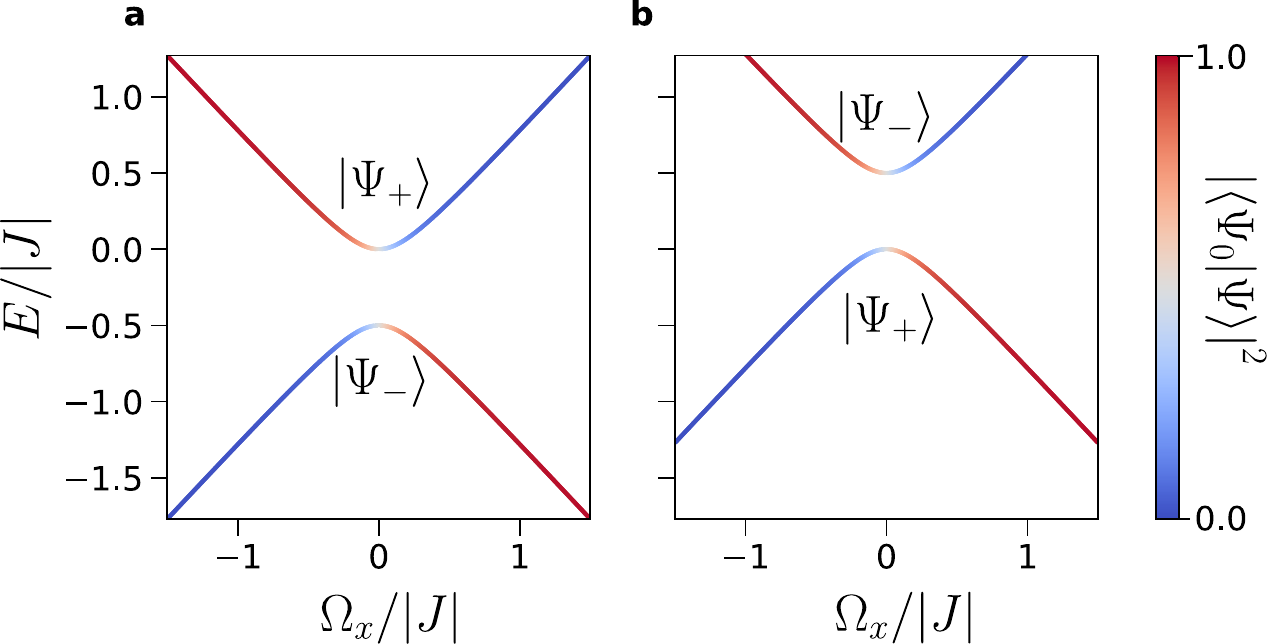}
    \caption{Spectrum of the positive parity subspace for a single pair of spins coupled by an XY model with a negative (a) or positive (b) coupling constant \(J\). The energy \(E\) is plotted against the applied field \(\Omega_x\) both in units of the coupling constant. The color code indicates the overlap with the state \(\ket{\Psi_0}\).}
    \label{fig:pairspectrum}
\end{figure}

In this product basis, the Hamiltonian reads
\begin{equation}
    H_0 = \begin{pmatrix}
        \frac{J}{4}-\Omega_x & 0 & 0 & -\frac{J}{4}\\
        0 & -\frac{J}{4} & \frac{J}{4} & 0 \\
        0 & \frac{J}{4} & -\frac{J}{4} & 0 \\
        -\frac{J}{4} & 0 & 0 & \frac{J}{4}+\Omega_x\\
    \end{pmatrix}.
\end{equation}
Here, we immediately recognize that independent of the field strength \(\Omega_x\), the Hamiltonian separates as a consequence of its spin flip symmetry \(\mathbb{Z}_2\) into two separate subspaces, of which only one subspace, namely the one spanned by the states \(\ket{+}_x\otimes\ket{+}_x\) and \(\ket{-}_x \otimes\ket{-}_x\), shows eigenstates depending on the annealing field \(\Omega_x\). Projected onto this subspace, the Hamiltonian reads
\begin{equation}
    H_0 = \begin{pmatrix}
        \frac{J}{4}-\Omega_x &  -\frac{J}{4}\\
        -\frac{J}{4} & \frac{J}{4}+\Omega_x\\
    \end{pmatrix} = \frac{J}{4} \mathbbm{1} +\begin{pmatrix}
        -\Omega_x &  -\frac{J}{4}\\
        -\frac{J}{4} & +\Omega_x\\
    \end{pmatrix},
\end{equation}
which takes, up to an energy shift, the same form as a Landau-Zener crossing with coupling \(-J/2\) and control parameter \(\Omega_x\).
In the case of zero magnetic field \(\Omega_x=0\), the eigenstates in the discussed subspace are
\begin{equation}
    \ket{\Psi_\pm} = \frac{\ket{+}_x\otimes\ket{+}_x \pm \ket{-}_x\otimes\ket{-}_x}{\sqrt{2}}
\end{equation}
with eigenenergies
\begin{equation}
    E_\pm = \frac{J}{4} \mp \frac{J}{4}.
\end{equation}
As a consequence, the energy of \(\ket{\Psi_+}\) is always zero, while the energy of \(\ket{\Psi_-}\) is \(J/2\), which is either positive or negative depending on the sign of \(J\). Thus, \(\ket{\Psi_-}\) is the ground state for \(J<0\), and \(\ket{\Psi_+}\) is the ground state of the system for \(J>0\). Upon application of a field, this ground state is adiabatically connected to the state \(\ket{\Psi_0} = \ket{+}_x\otimes\ket{+}_x\), as shown in Fig. \ref{fig:pairspectrum}.

\subsubsection{ZFA and FA for a single pair}
Within this spectral picture, both FA and ZFA susceptibilities can be interpreted. For very fast ramps, in both cases the final state at the end of the ramp is \(\ket{\Psi_0}\), which only oscillates in the \(x-z-\)plane upon an applied field in the \(y-\)direction, but shows no response. This explains why, for fast ramps, both the ZFA and FA protocols yield no magnetic response.

In the limit of an adiabatic ramp, we start our discussion with the ZFA protocol. In this case, for \(J<0\) the system is prepared in the state \(\ket{\Psi_-}\) and for \(J>0\) in the state \(\ket{\Psi_+}\). These states themselves show no \(y-\)magnetization, i.e.
\begin{equation}
    \expval{M_y}{\Psi_\pm}=0,
\end{equation}
where
\begin{equation}
    M_y = \frac{S_y^{(1)}+S_y^{(2)}}{2}
\end{equation}
is the operator of the average \(y-\)magnetization.
However, as in the ZFA protocol a dc linear susceptibility is measured, we are interested how the magnetization of these states changes as a function of an applied infinitesimal probe field. If the eigenstates \(\ket{m}\) of \(H_0\) are perturbed by an infinitesimal probe field \(\Omega_y\) in \(y-\)direction, we can obtain the adiabatically connected eigenstates in first order non-degenerate perturbation theory
\begin{equation}
    \ket{\Tilde{m}} =\ket{m} - \Omega_y \sum_{n\neq m} \frac{ \mel{n}{S_y^{(1)}+S_y^{(2)}}{m}}{E_m - E_n} \ket{n}.
\end{equation}
These states thus show the magnetization
\begin{equation}
    \expval{M_y}{\Tilde{m}} = \expval{M_y}{m} + 4 \Omega_y \sum_{n\neq m} \frac{\left|\mel{m}{M_y}{n}\right|^2}{E_n-E_m} +\mathcal{O}\left(\Omega_y^2\right),
    \label{eq:vanFleck_suscept}
\end{equation}
from which we can read off the magnetic susceptibility as
\begin{equation}
    \chi= 4 \sum_{n\neq m} \frac{\left|\mel{m}{M_y}{n}\right|^2}{E_n-E_m}.
\end{equation}
In order to calculate these susceptibilities, we also need the eigenenergies and eigenstates of the negative parity subspace, spanned by the states \(\ket{+}_x\otimes\ket{-}_x\) and \(\ket{-}_x\otimes\ket{+}_x\). In this subspace, the eigenstates are given as
\begin{equation}
    \ket{\Phi_\pm} = \frac{\ket{+}_x\otimes\ket{-}_x \pm \ket{-}_x\otimes\ket{+}_x}{\sqrt{2}},
\end{equation}
with respective eigenenergies 
\begin{equation}
    E^\Phi_\pm = -\frac{J}{4} \pm \frac{J}{4}.
\end{equation}
We first observe that \(\ket{\Psi_+}\) and \(\ket{\Phi_+}\) are degenerate with energy \(E=0\). However, their magnetic coupling by a \(y-\)field also vanishes
\begin{equation}
    \mel{\Phi_+}{M_y}{\Psi_+}=0,
\end{equation}
such that we may apply non-degenerate perturbation theory in that basis.
In summary, the couplings between different eigenstates read
\begin{align}
   \mel{\Psi_+}{M_y}{\Psi_-} &=-\frac{i}{2} \braket{\Psi_+}{\Phi_+} = 0, \\
   \mel{\Psi_+}{M_y}{\Phi_+} &=\frac{i}{2} \braket{\Psi_+}{\Psi_-} = 0,\\
   \mel{\Psi_+}{M_y}{\Phi_-} &=\bra{\Psi_+} \cdot 0 = 0, \\
   \mel{\Psi_-}{M_y}{\Phi_+} &=\frac{i}{2} \braket{\Psi_-}{\Psi_-} = \frac{i}{2}, \\
   \mel{\Psi_-}{M_y}{\Phi_-} &=\bra{\Psi_-} \cdot 0 = 0. 
\end{align}
Plugging this back into Eq. \ref{eq:vanFleck_suscept}, we obtain for the susceptibility of state \(\ket{\Psi_+}\) in the ZFA protocol
\begin{equation}
    \chi_\mathrm{ZFA}^{\ket{\Psi_+}} = 0,
\end{equation}
and for that of \(\ket{\Psi_-}\)
\begin{equation}
    \chi_\mathrm{ZFA}^{\ket{\Psi_-}} = 4 \cdot \frac{\left|i/2\right|^2}{0-J/2}=-\frac{2}{J}.
\end{equation}
When \(J<0\) and \(\ket{\Psi_-}\) is the ground state, the excited state \(\ket{\Psi_+}\) shows no response, while the ground state shows a positive response. When \(J>0\) and \(\ket{\Psi_+}\) is the ground state, the excited state shows a negative response while the ground state shows no response. In either case, the ground state always shows a larger magnetic response than the excited state.

In the case of the FA protocol, we treat the sum of the annealing field \(\Omega_x(t)\) and the constant probe field \(\Omega_y\) as a single field which decreases in amplitude approximately linear in time and slowly rotates from the \(x-\) to the \(y-\)direction. We can choose a time-dependent basis \(\ket{+}_x(t)\) and \(\ket{-}_x(t)\) such that the rotating field always points in the \(x-\)direction. Due to the \(U(1)\) symmetry of the XY-Hamiltonian, the spin-spin interaction is not affected by this rotation. In the limit of an adiabatic ramp, the annealing ramp prepares the same states as in the ZFA protocol, except that now the probe field does not point in the \(y-\) but in the \(x-\)direction in the time-dependent basis at the end of the annealing ramp. Using the same reasoning as in the derivation of Eq. \ref{eq:vanFleck_suscept}, we obtain for a response for a probe field in the \(x-\)direction
\begin{equation}
    \chi= 4 \sum_{n\neq m} \frac{\left|\mel{m}{M_x}{n}\right|^2}{E_n-E_m}.
    \label{eq:vanFleck_FA}
\end{equation}
To use this formula for the magnetic response of each eigenstate, we calculate the couplings between different eigenstates
\begin{align}
   \mel{\Psi_+}{M_x}{\Psi_-} &= \frac{1}{2} \braket{\Psi_+}{\Psi_+} = \frac{1}{2}, \\
   \mel{\Psi_+}{M_x}{\Phi_+} &=\bra{\Psi_+} \cdot 0 = 0,\\
   \mel{\Psi_+}{M_x}{\Phi_-} &=\bra{\Psi_+} \cdot 0 = 0, \\
   \mel{\Psi_-}{M_x}{\Phi_+} &=\bra{\Psi_-} \cdot 0 = 0, \\
   \mel{\Psi_-}{M_x}{\Phi_-} &=\bra{\Psi_-} \cdot 0 = 0. 
\end{align}
Plugging this into Eq. \ref{eq:vanFleck_FA}, we find for the susceptibility of the state \(\ket{\Psi_+}\) in the FA protocol
\begin{equation}
    \chi_\mathrm{FA}^{\ket{\Psi_+}} = 4 \cdot \frac{\left|1/2\right|^2}{J/2-0}=\frac{2}{J},
\end{equation}
and of the state \(\ket{\Psi_-}\)
\begin{equation}
    \chi_\mathrm{FA}^{\ket{\Psi_-}} = 4 \cdot \frac{\left|1/2\right|^2}{0-J/2}=-\frac{2}{J},
\end{equation}
Again, independent of the sign of \(J\), we find that the ground state always shows a larger magnetic response than the excited state.

These calculations enable a test whether a bifurcation between ZFA and FA magnetizations occurs for a single pair of spins. When the coupling \(J<0\) is negative and \(\ket{\Psi_-}\) is the annealed ground state, both ZFA and FA show the same response \(\chi=-2/J\) and no energetic-magnetic hysteresis occurs. On the other hand, when \(J>0\) and \(\ket{\Psi_+}\) is the annealed ground state, the difference in the susceptibilities is
\begin{equation}
    \chi_\mathrm{FA}^{\ket{\Psi_+}}-\chi_\mathrm{ZFA}^{\ket{\Psi_+}} = \frac{2}{J}-0 = \frac{2}{J}>0.
    \label{eq:hysteresis_theory}
\end{equation}
In a single pair the FA susceptibility is thus always larger or equal to the ZFA susceptibility.

For finite ramp times and large initial fields \(\Omega_x^0\), the annealing ramp can be approximated as half a Landau-Zener crossing, where a superposition state between \(\ket{\Psi_+}\) and \(\ket{\Psi_-}\) is realized. The overlap of this superposition state with the ground state may be approximated following Ref. \cite{damski_adiabatic_2006} using the adiabatic impulse approximation as
\begin{equation}
    p \approx \frac{1}{2} + \frac{1}{2}\sqrt{1-\frac{2}{b^2+b\sqrt{b^2+4}+2}},
    \label{eq:AI_p(b)}
\end{equation}
which increases monotonically with the dimensionless ramp time
\begin{equation}
    b = \frac{\pi J^2 t_r}{32 \Omega_x^0}.
\end{equation}
Therefore, we observe that with increasing ramp time \(t_r\) and thus increasing \(b\), the system is more likely to stay in the ground state, where the magnetic response is largest. On the other hand, when \(t_r\) and thus \(b\) decrease, the prepared state of the annealing ramp has more overlap with the excited state that shows a smaller response. Therefore, for a single pair we expect that the both the ZFA and FA susceptibilities increase monotonically with \(t_r\) in the limit of small and intermediate ramp times where the adiabatic impulse approximation applies.

\subsubsection{ZFA and FA for an ensemble of pairs}

To treat an ensemble of pairs at a given ramp time \(t_r\) and initial field \(\Omega_x^0\), we need to estimate whether a pair interacting with a coupling \(J\) from the ensemble can be treated diabatically showing no response or adiabatically showing the response derived above. To that end we solve Eq. \ref{eq:AI_p(b)} for the ramp time \(t_r\) at which the system stays in the ground state with a probability \(p\) as
\begin{equation}
    t_r = \frac{16 \Omega_x^0}{\pi J^2} \frac{\left(2p-1\right)^2}{\sqrt{p\left(1-p\right)}} \sim \frac{1}{J^2}, \quad 1/2 \le p \le 1.
\end{equation}
Defining a threshold value \(p_\mathrm{th} \lesssim 1\) such that we treat the pair approximately as completely diabatically if \(p<p_\mathrm{th}\) and perfectly adiabatically for \(p>p_\mathrm{th}\), a threshold coupling \(J_\mathrm{th}\) can be defined for a given ramp time \(t_r\) and initial field \(\Omega_x\) as
\begin{equation}
    J_\mathrm{th} = \sqrt{\frac{16 \Omega_x^0}{\pi t_r} \frac{\left(2p_\mathrm{th}-1\right)^2}{\sqrt{p_\mathrm{th}\left(1-p_\mathrm{th}\right)}}} \sim \frac{1}{\sqrt{t_r}}
\end{equation}
Within this picture, pairs with coupling strengths \(\left|J\right|>J_\mathrm{th}\) are assumed to evolve adiabatically and thus contribute to the response according to the expressions derived above. Conversely, pairs with \(\left|J\right|<J_\mathrm{th}\) are assumed to evolve diabatically, since the ramp is too fast compared to the coupling strength, yielding zero response.

\begin{figure}[!ht]
    \centering
    \includegraphics[width=0.9\linewidth]{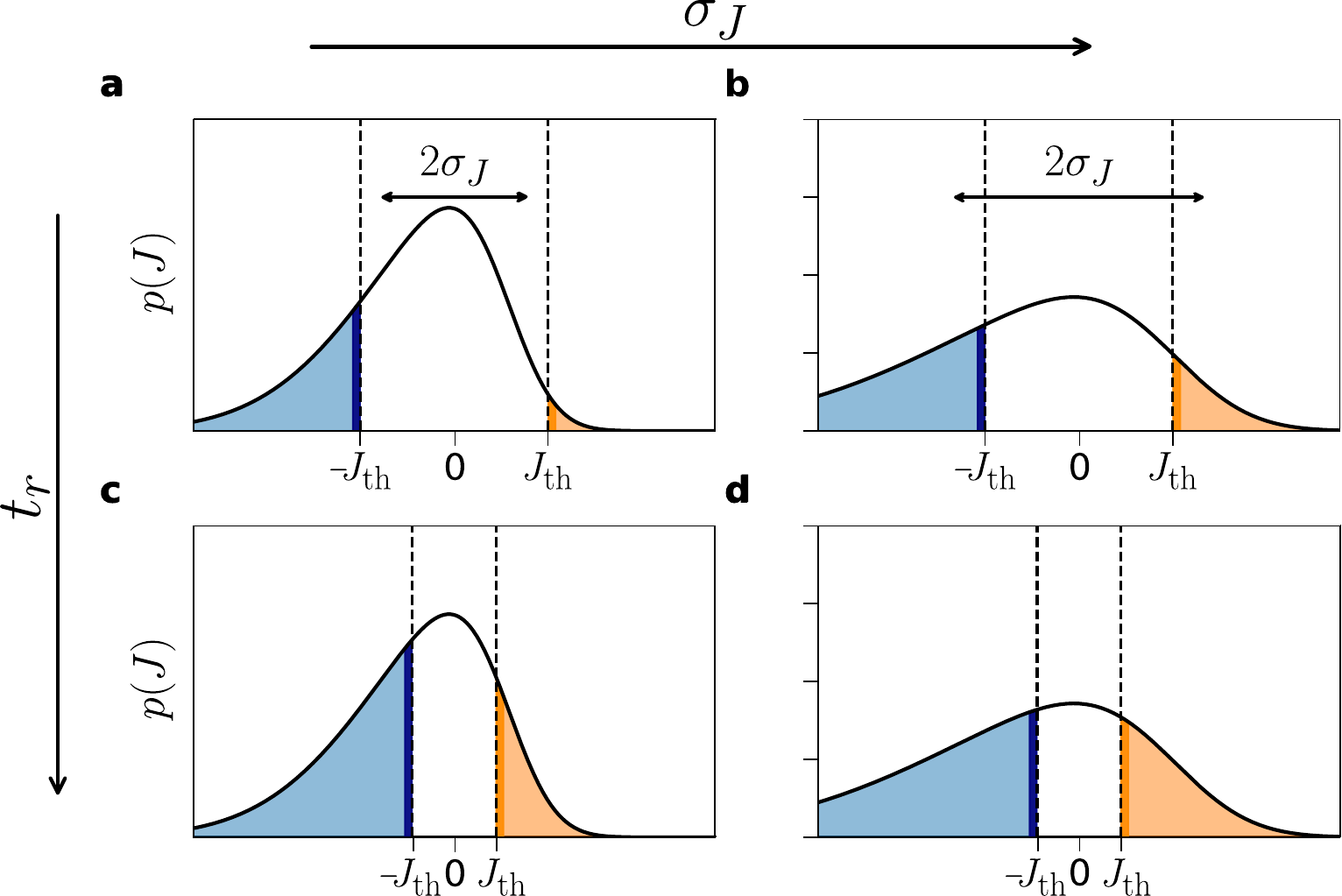}
    \caption{Coupling distributions \(p(J)\) as a function of the couplings strength \(J/J_\mathrm{med}\). Black dotted vertical lines denote the threshold coupling strength \(J_\mathrm{th}(t_r)\), which is large for the top row plots a and b (small \(t_r\)) and small for the bottom row plots c and d (large \(t_r\)). The way how \(t_r\) changes is indicated by a black arrow left of the figure. The coupling distributions can be characterized by a width \(\sigma_J\), indicated by a black arrow in the top row plots. The distribution is narrower in the left column plots a and c (weak disorder configuration) and broader in the right column plots b and d (strong disorder configuration). Pairs with positive and negative couplings which contribute to the magnetic response are shown in orange and blue, respectively. The pairs with strongest magnetic response are highlighted in dark orange and dark blue, respectively.}
    \label{fig:coupling_distribution_eff}
\end{figure}

As discussed in Sec. \ref{sec:distribution}, the distribution of nearest-neighbor couplings in a positionally disordered dipolar interacting system is peaked around \(J\approx0\) and decays toward the tails. Its width can be characterized by the standard deviation \(\sigma_J\). Combining this coupling distribution with an adiabatic and diabatic treatment of the ZFA and FA susceptibilities allows us to qualitatively explain the features observed in both numerical simulations and experiments, as illustrated in Fig. \ref{fig:coupling_distribution_eff}.

In Fig. \ref{fig:coupling_distribution_eff}a we show a couplings distribution, where for a given \(\Omega_x^0\) and \(t_r\) only the couplings with \(\left|J\right|>J_\mathrm{th}\) contribute to the magnetic response.
These couplings are highlighted in blue for negative and in organge for positive couplings, respectively. The magnetic response is zero in the case of \(J>0\) for the ZFA protocol and \(2/\left|J\right|\) for \(J<0\) in the ZFA and for both positive and negative couplings in the FA protocol. 
As the response is inversely proportional to the coupling strength, the magnetic response is dominated by the weakest interacting pairs which are still adiabatically prepared in the ground state of the XY pair Hamiltonian. 
These are exactly the pairs with \(\left|J\right|\approx J_\mathrm{th}\), which are highlighted in dark orange and dark blue for positive and negative \(J\), respectively. Hysteresis occurs only for positive \(J\). As the difference between the ZFA and FA susceptibilities is inversely proportional to \(J\) according to Eq. \ref{eq:hysteresis_theory}, this difference is dominated by pairs with coupling strength \(J \gtrsim J_\mathrm{th}\), i.e. the couplings highlighted in dark orange.

When the disorder is increased while all other parameters are kept fixed, the coupling distribution broadens, whereas the threshold coupling \(J_\mathrm{th}\) remains unchanged. This situation is illustrated in Fig. \ref{fig:coupling_distribution_eff}b. The broader distribution carries more weight in its tails, resulting in a larger fraction of pairs with \(\left|J\right|>J_\mathrm{th}\). Consequently, more pairs are adiabatically prepared in the ground state of the XY Hamiltonian and therefore contribute to the magnetic response. As a result, increasing the disorder broadens the coupling distribution and enhances the magnetic susceptibility in both the ZFA and FA protocols. Furthermore, the increased weight for \(J>J_\mathrm{th}\) leads to a more pronounced magnetic hysteresis in a more strongly disordered system.

The second effect arises when the ramp time \(t_r\) is increased, or equivalently when the inverse ramp time \(1/t_r\) is decreased, as illustrated in Fig.~\ref{fig:coupling_distribution_eff}c. Since \(J_\mathrm{th} \sim t_r^{-1/2}\), the threshold coupling \(J_\mathrm{th}\) is reduced compared to the situation shown in Fig.~\ref{fig:coupling_distribution_eff}a, while the coupling distribution remains unchanged. Consequently, a larger fraction of pairs satisfies the condition \(\left|J\right|>J_\mathrm{th}\). Physically, pairs that evolved diabatically and therefore did not contribute to the magnetic response for the shorter ramp time can now follow the ramp adiabatically and thus contribute to the response. As a result, the increased fraction of pairs with \(\left|J\right|>J_\mathrm{th}\) enhances both the magnetic susceptibility and the energetic-magnetic hysteresis, following the same reasoning as for the broader coupling distribution shown in Fig.~\ref{fig:coupling_distribution_eff}b.

The strongest hysteresis and magnetic response are expected in the limit of strong disorder at large \(t_r\), where the combination of a broad coupling distribution and a small \(J_\mathrm{th}\) causes nearly all pairs to contribute to the magnetic response. This situation is illustrated in Fig. \ref{fig:coupling_distribution_eff}d.

\section{Experimental details}

\subsection{Preparation of the Rydberg atom distribution}
We load \ce{^{87}Rb} atoms in a \SI{780}{\nano\meter} 2D MOT using the D2 line between the states \(\ket{g}=\ket{5S_{1/2},F=2,m_F=2}\) and \(\ket{e}=\ket{5P_{3/2},F=3,m_F=3}\). A pusher beam pushes the atoms into the science chamber, where about \num{100000} atoms are collected in a \SI{780}{nm} 3D MOT. 
From there, we load \(N_g\) atoms into a \SI{9}{\degree} crossed dipole trap at \SI{1064}{\nano\meter} at \SI{11}{\micro\kelvin} and increase the magnetic field to its final value of \SI{185}{\gauss}. The size of the trap can be changed by letting the atoms fly freely during a time \(t_\mathrm{tof}\). The spatial extent of the optical dipole trap can be approximated by a Gaussian with characteristic widths \(\sigma_x\) and \(\sigma_y=\sigma_z\). 
We use a 2-photon excitation scheme with a \SI{780}{\nano\meter} coupling beam, which is \SI{97}{\mega\hertz} blue detuned from the \(\ket{g}\rightarrow\ket{e}\) transition, and a \SI{480}{\nano\meter} probe beam, to excite the atoms in the 3D MOT from \(\ket{g}\) to the Rydberg state \(\ket{r1}=\ket{61S_{1/2},m_J=1/2}\). 
During an excitation time \(t_\mathrm{exc}\), \(N_r\) atoms are excited into the Rydberg state from a thermal cloud confined in an optical dipole trap. Their positions inherit the cloud's non-uniform spatial profile, while the Rydberg blockade suppresses close pair separations. We characterize the resulting spatial distribution of the Rydberg atoms by the mean nearest-neighbor distance \(r_\mathrm{mean}\) and the blockade radius \(r_\mathrm{bl}\), which sets the characteristic minimum separation between them.  Microwave radiation at frequency \(\omega\), generated by a Keysight M8195A arbitrary waveform generator, is used to couple the Rydberg state \(\ket{r1}\) to a second Rydberg state \(\ket{r2}\). These two states form an effective 2-level system, which is the basis for the presented experiments. The experimental values for the quantities in both the weakly and strongly disordered regime are summarized in table \ref{tab:expvals}.
\begin{table*}[!htb]
    \centering
    \begin{tabular}{c c c c c c c c c c c c}
    \toprule
    & \(N_g\)   & \(N_s\) & \(t_\mathrm{exc}\)[\unit{\micro\second}] &\(t_\mathrm{tof}\)[\unit{\micro\second}] & \(\sigma_x\)[\unit{\micro\meter}] & \(\sigma_y,\sigma_z\)[\unit{\micro\meter}] &\(r_\mathrm{mean}\)[\unit{\micro\meter}] &\(r_\mathrm{bl}\)[\unit{\micro\meter}] &\(\ket{r2}\) &\(\omega/(2\pi)[\unit{\mega\hertz}]\)  \\
    \midrule
    weak disorder &10300 & 2250 & \num{5} &\num{600} &\num{140} &\num{49} & \num{18.7} &\num{8.15} &\(\ket{61P_{1/2},m_J=1/2}\) &  \num{15791.2} \\
    strong disorder &9300 &2100 &\num{1} & \num{2000} &\num{137} &\num{97} &\num{26.5} &\num{7.85} &\(\ket{61P_{3/2},m_J=1/2}\) & \num{16324.4}\\
    \bottomrule
    \end{tabular}
    \caption{Comparison of experimental parameters in the two discussed disorder configurations.}
    \label{tab:expvals}
\end{table*}

\subsection{Comparison of couplings distributions between numerical simulation and experiment}
\label{sec:distribution}
Although the spin configurations used in the few-particle simulations are motivated by a Rydberg excitation process, the resulting coupling distributions differ quantitatively from those realized experimentally.

In the twelve-particle simulations, the spin positions are sampled uniformly within a three-dimensional unit sphere. Inspired by the Rydberg blockade, configurations are post-selected to enforce a minimum interspin distance \( r_\mathrm{bl} < 1 \), which sets an upper bound on the interaction strength. The disorder strength is controlled by varying \( r_\mathrm{bl} \): we use \( r_\mathrm{bl} = 0.39 \) for the weakly disordered regime and a smaller blockade of \( r_\mathrm{bl} = 0.12 \) for the strongly disordered regime.

\begin{figure}[!t]
    \centering
    \includegraphics[]{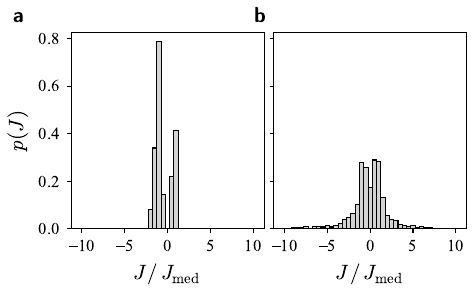}
    \caption{Nearest-neighbor coupling distribution used for exact diagonalization in the weak (a) and strong (b) disorder regime. The binwidth is $J_\mathrm{med}/2$.}
    \label{fig:couplings_dist_num}
\end{figure}

For each spin \( i \), we define its nearest neighbor as \( j(i) = \underset{j}{\arg\max}\, |J_{ij}| \) and the corresponding nearest neighbor coupling as \( J(i) = J_{i j(i)} \), where the couplings follow the dipolar form discussed in the main text. For each of the 100 (250) disorder realizations in the weak (strong) disorder case, the full set of \( J(i) \) is sampled and, after normalization by the median interaction strength of that realization, combined into a histogram over all realizations. Fig. ~\ref{fig:couplings_dist_num} shows the resulting normalized histogram, which approximates the coupling distribution \( p(J) \) in both regimes. 

We observe substantial frustration, evidenced by the presence of both positive and negative couplings with comparable probabilities within each disorder realization. Negative couplings occur slightly more frequently, reflecting the sign change of the angular factor
\( 1 - 3\cos^2\theta \), which depends on the orientation of the interatomic axis relative to the quantization axis. Furthermore, the weakly disordered configuration exhibits a narrower coupling distribution than the strongly disordered one. This reflects the spatial correlations induced by \( r_\mathrm{bl} \), which render the couplings more uniform in magnitude. Owing to the finite system size of 12 particles, nearest-neighbor distances that differ by an order of magnitude are rare;
consequently, a suppression of the distribution is observed for \( J \ll J_\mathrm{med} \).

\begin{figure}[!b]
    \centering
    \includegraphics[width=\linewidth]{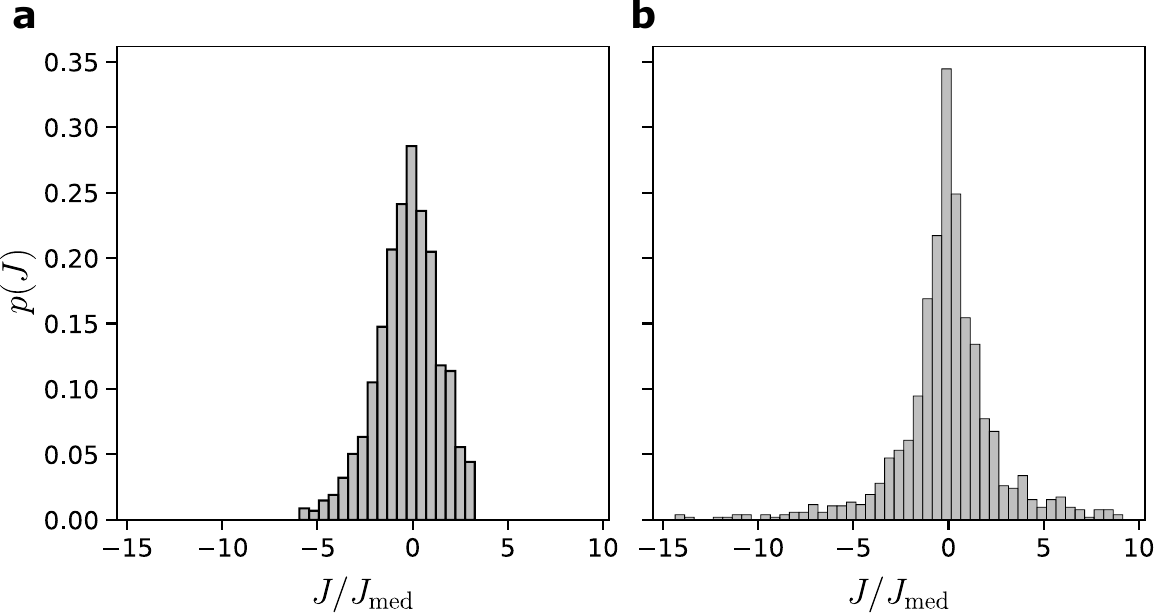}
    \caption{Simulation of the nearest-neighbor coupling distribution in the experiment for weak (a) and strong (b) disorder, normalized by \(J_\mathrm{med}\). The binwidth is by choice \(J_\mathrm{med}/2\). The normalized histogram provides an estimate of the coupling probability distribution \(p(J)\).}
    \label{fig:couplings_distro_exp}
\end{figure}
In the experiment, atoms encoding spin states are excited to Rydberg levels from an atomic cloud that has freely expanded from a crossed-beam dipole trap and retains a Gaussian density profile.
To estimate the resulting nearest-neighbor coupling distribution, we exploit self-averaging: for sufficiently large samples, spatial averaging is equivalent to a disorder average, as commonly assumed in spin-glass simulations \cite{dahlberg_spin-glass_2024}.

From absorption images of the ground-state atoms, we infer a Gaussian spatial distribution.
Following the procedure of Ref.~\cite{signoles_glassy_2021}, and including the spatial profiles of the excitation lasers as well as the excitation duration, we numerically generate disorder realizations of Rydberg atom positions.
To reduce numerical complexity, we assume an isotropic \( C_6 \) coefficient and thus an isotropic blockade, using the median value over all directions.
As a consequence, the sampled disorder realizations do not accurately describe regimes in which the mean interparticle distance approaches the blockade radius.
However, in both experimentally realized disorder configurations, the median interparticle distance exceeds twice the median blockade radius.
Close particle pairs therefore occur rarely, and we expect the simulations to faithfully capture the coupling distributions.
For each realization, the couplings are computed using the coefficient \( C_3 \) obtained from the \textit{PairInteraction} package \cite{weber_calculation_2017, mogerle_pairinteraction_2026}.

Fig. ~\ref{fig:couplings_distro_exp} shows the normalized histogram of nearest-neighbor couplings \( J \) obtained from a single disorder realization, providing an estimate of the distribution \( p(J) \).
As in the numerical results for the system consisting of 12 spins, negative couplings occur with slightly higher probability due to the angular dependence of the interactions and the trap geometry.
The weakly disordered configuration again exhibits a narrower distribution, reflecting spatial correlations induced by the blockade radius, which render the median and maximal coupling strengths more similar.

In contrast to the twelve-particle numerical simulations, the experimental system contains thousands of spins and, therefore, exhibits many nearest-neighbor couplings weaker than the median interaction strength.
This produces a pronounced peak near \( J \approx 0 \), which is absent in the simulations for the 12 particles. In addition, the experiment samples a range of densities because the ground-state atomic distribution is Gaussian, whereas the numerical simulations assume a uniform density. This density inhomogeneity further enhances the peak near \( J \approx 0 \), as atoms near the edges of the Gaussian distribution experience very weak nearest-neighbor coupling.
Nevertheless, the central qualitative feature of the coupling distributions, namely the presence of long tails toward large couplings in the strongly disordered regime and their absence in the weakly disordered regime, is observed both in the numerical simulations and in the experiment.

According to the pair model discussed in Sec.~\ref{sec:pairpicture}, the enhanced tails in the coupling distributions toward stronger disorder account for the increase in both the ZFA and FA susceptibilities at low energies, as well as for the more pronounced bifurcation at low energies and stronger disorder. These effects explain the qualitative agreement between the numerical simulations and the experiment.
In the weak-disorder regime, the experiment averages over regions with different disorder strengths across the Gaussian shaped cloud of atoms in their ground state. This averaging produces longer tails in the coupling distribution than those observed in the numerical simulations. It may therefore explain the small low-energy bifurcation seen in the experiment but absent from the numerics, and may partially account for the quantitative differences between the experimental and numerical values.

\subsection{Engineering the effective magnetic field with an arbitrary waveform generator}

The microwave radiation created by the aforementioned AWG is guided to air using a Horn antenna. The polarization is cleaned up to vertical direction, parallel to the magnetic field, using a wire grid polarizer. We define the axis along the magnetic field to be the quantization axis, which coincides with the \(z-\)axis. As a consequence, the Hamiltonian for the interaction of the atom with the microwave radiation at frequency \(\omega\) reads
\begin{align*}
    H(t) &= -E_z(t)\hat{d}_z +H_0\\
    &= -(i\frac{E}{2}e^{i\phi}e^{-i\omega t}-i\frac{E}{2}e^{-i\phi}e^{i\omega t})\hat{d}_z+H_0\\
\end{align*}
where \(E\) defines the amplitude and \(\phi\) the phase of the microwave drive, \(\hat{d}\) the electric dipole operator of the atom and \(H_0\) the atomic Hamiltonian. Assuming a dipole allowed transition between two atomic states \(\ket{\uparrow}\) and \(\ket{\downarrow}\), we can write the Hamiltonian in the basis of these two states
and apply the rotating wave approximation to this Hamiltonian
\begin{align*}
    H &= \begin{pmatrix}
        0 & -i\frac{E}{2}e^{-i\phi}d\\
        i\frac{E}{2}e^{i\phi}d & 0
    \end{pmatrix}
    = \frac{1}{2}\begin{pmatrix}
        0 & -i\Omega^\ast\\
        i\Omega & 0
    \end{pmatrix}\\
    &= \frac{1}{2} \begin{pmatrix}
        0 & -i\Re(\Omega)-\Im(\Omega)\\
        i\Re(\Omega)-\Im(\Omega) &0
    \end{pmatrix}\\
    &=  -\Im(\Omega)\frac{1}{2} \begin{pmatrix}
        0 & 1\\
        1 &0
    \end{pmatrix}
    + \Re(\Omega)\frac{1}{2}  \begin{pmatrix}
        0 & -i\\
        i &0\\
    \end{pmatrix}\\
    &= -\Im(\Omega)\hat{S}_x+\Re(\Omega)\hat{S}_y
\end{align*}
where \(d=\mel{\uparrow}{\hat{d}}{\downarrow}\in \mathbb{R}\) is the dipole matrix element between the two atomic states, and \(\Omega = dEe^{i\phi} \) the Rabi frequency of the driven transition.
Using these conventions, we immediately see that the phase \(\phi\) of the drive \(E_z(t)=-E\sin(\omega t+\phi)\) is exactly equal to the phase of the Rabi frequency. Identifying \(\Im(\Omega)\coloneqq\Omega_x\) and \(-\Re(\Omega)\coloneqq\Omega_y\) we recover the external field Hamiltonian 
\begin{equation}
    H_\mathrm{field}=-\mathbf{\Omega}\hat{\mathbf{S}}
\end{equation}
from the main text, setting \(\Omega_z=0\).

Experimentally, the time-dependent electric field is created using a Keysight M8195A arbitrary waveform generator with an analog bandwith of \SI{25}{\giga\hertz} and a control of the waveform at \SI{65}{\giga\sample\per\second}. As a consequence, radiation of up to 25GHz can be both phase and amplitude modulated on a \SI{16}{\pico\second} timescale. By varying both phase and amplitude of the drive as a function of time, we can also vary phase and amplitude of the Rabi frequency.

Rabi pulses with Rabi frequency \(\Omega\) are used for state initialization. In the response measurement protocol, the ZFA and FA magnetizations for \(t_{r}=\SI{3}{\micro\second}\) and \(t_{w}=\SI{1}{\micro\second}\) are measured as a function of the probe field \(-\Omega_y^\mathrm{max} \leq \Omega_y \leq \Omega_y^\mathrm{max}\). This measurement protocol is used to estimate the linear regime.
In the hysteresis measurement protocol, the linear susceptibility \(\chi\) is measured as a function of the ramp speed. The Rabi frequency \(\Omega\) is kept at a constant but lower level than in the response measurement protocol, which allows to achieve slower ramp speeds at constant \(J_\mathrm{med}\) and \(t_{r}\). Ramp speed is varied by changing \(t_{r}\) in an interval from \SI{10}{\nano\second} to \SI{10}{\micro\second}. For each ramp speed, the susceptibility is extracted by varying the probe field in the interval \(-\Omega_y^\mathrm{max} \leq \Omega_y \leq \Omega_y^\mathrm{max}\), where the response is approximately linear, and performing a linear fit. The experimental values for all Rabi frequencies are listed in table \ref{tab:rabi}
\begin{table}[!htpb]
    \centering
    \begin{tabular}{c c c c c}
       {\thead[b]{disorder\\regime}}  & {\thead[b]{measurement\\protocol}} & {\thead[b]{\(\Omega/(2\pi)\) \\ \([\unit{\mega\hertz}]\)}} & {\thead[b]{\(\Omega_{x}/(2\pi)\)\\ \([\unit{\mega\hertz}]\)}}  & {\thead[b]{\(\Omega_{y}^\mathrm{max}/(2\pi)\)\\ \([\unit{\mega\hertz}]\)}}  \\
       \toprule
       weak & response & \num{12.18} & \num{3.65} & \num{4.87}\\
       weak & hysteresis & \num{11.75} & \num{3.53} & \num{0.47}\\
       \midrule
       strong & response & \num{16.3} & \num{4.89} & \num{4.89}\\
       strong & hysteresis&\num{16.01}&\num{4.80}& \num{0.64}\\
       \bottomrule
    \end{tabular}
    \caption{Experimental values for Rabi frequency amplitudes of \(\pi/2-\)pulses \(\Omega\), annealing ramp \(\Omega_{x}\), and maximal probe field \(\Omega_{y}^\mathrm{max}\) used for the different measurements in different regimes.}
    \label{tab:rabi}
\end{table}

\subsection{Calculation of dispersion coefficients in strong magnetic fields}
Without an external electric or magnetic field, the electronic eigenstates in alkaline atoms are the eigenstates of the bare atomic Hamiltonian \(H_0\) and are labeled by their principal quantum number \(n\), their orbital angular momentum quantum number \(L\), their total angular momentum quantum number \(J=L+S\) and their magnetic quantum number \(m_J\). If a magnetic field is applied, to first approximation a linear Zeeman shift applies, and the wavefunctions of the eigenstates do not change. For Rubidium at \SI{185}{\gauss}, this approximation no longer holds. As a consequence, we diagonalize the atomic Hamiltonian in a magnetic field \(\vb{B}=B\hat{e}_z\)
\begin{equation}
    H_a = H_0 +\mu_B\left(g_L\vb{L}+g_S\vb{S}\right)\vb{B}+\frac{\left(\vb{d}\times\vb{B}\right)^2}{8m_e}
\end{equation}
where \(\mu_B\) is the Bohr magneton, \(\vb{L}\) the orbit angular momentum operator, \(\vb{S}\) the spin operator, \(g_S\) and \(g_L\) the corresponding Landé g-factors, \(m_e\) the electron mass and \(\vb{d}\) the electric dipole operator. The \(z\)-axis is chosen as the quantization axis.
The terms linear and quadratic in the magnetic field will mix states with the same magnetic quantum number and orbital angular momentum quantum numbers of the same parity, and the eigenstates of \(H_a\) will hence be a superposition of such states.
Relevant for the states in the main paper is especially the strong admixing between the states \(\ket{61P_{1/2},m_J={1/2}}\) and \(\ket{61P_{3/2},m_J={1/2}}\).
We want to study the influence of this admixture on the dispersion coeffecients, which can be obtained by first and second-order perturbation theory.
The total Hamiltonian of two atoms interacting via dipole-dipole interactions is given as
\begin{equation}
    H =H_a\otimes\mathbb{I}+\mathbb{I}\otimes H_a + H_{\rm{DDI}}
\end{equation}
where the dipole-dipole Hamiltonian is given as
\begin{align}
\hat{H}_{\rm{DDI}} = \frac{1}{4\pi \epsilon_0} \Bigg[ \frac{1-3 \cos^2 \theta}{2r^3} \left( 2\hat{d}_1^{\,0}\hat{d}_2^{\,0} + \hat{d}_1^+ \hat{d}_2^- + \hat{d}_1^- \hat{d}_2^+ \right) \nonumber \\
-  \frac{3 \sin \theta \cos \theta}{\sqrt{2}r^3}\left(( \hat{d}_1^+ \hat{d}_2^{\,0}+ \hat{d}_1^{\,0} \hat{d}_2^+)e^{-i \phi} - (\hat{d}_1^{\,0} \hat{d}_2^- + \hat{d}_1^- \hat{d}_2^{\,0}) e^{i \phi}  \right) \nonumber  \\
- \frac{3 \sin^2 \theta}{2r^3} \left(  \hat{d}_1^+ \hat{d}_2^+ e^{-2 i \phi} +  \hat{d}_1^- \hat{d}_2^- e^{2i \phi} \right) \Bigg] \quad. \label{eq:HDD}
\end{align}
where \(d_i^0\) and \(d_i^{\pm1}\) are the components of the electric dipole operator acting on atom \(i\) in the spherical basis. The basis of the coordinate system is chosen such that the first atoms is in the origin of the coordinate system, and the second atoms has the coordinates \(\left(r,\theta,\phi\right)\) in spherical coordinates. For an atom in the \(\ket{nS}\) and an atom in the \(\ket{nP}\) states, the two-atom states \(\ket{nS,nP}\) and \(\ket{nP,nS}\) are degenerate. If we set the energy level of these two states to zero by gauge freedom, the Hamiltonian in this basis reads
\begin{align}
    \hat{H}&=\frac{1-3\cos(\theta)^2}{r^3} \frac{1}{4\pi\epsilon_0}\frac{2\hat{d}_1^{\,0}\hat{d}_2^{\,0} + \hat{d}_1^+ \hat{d}_2^- + \hat{d}_1^- \hat{d}_2^+}{2}\\
    &= \frac{1-3\cos(\theta)^2}{r^3} \underbrace{\frac{1}{4\pi\epsilon_0}\frac{\expval{2\hat{d}_1^{\,0}\hat{d}_2^{\,0} + \hat{d}_1^+ \hat{d}_2^- + \hat{d}_1^- \hat{d}_2^+}}{2}}_{C_3} \begin{pmatrix}0&1\\1&0\end{pmatrix}\\
    &=\frac{1-3\cos(\theta)^2}{r^3}C_3\begin{pmatrix}0&1\\1&0\end{pmatrix}
\end{align}
Due to the eigenstates of \(H_a\) being a superposition of the states \(\ket{61P_{1/2},m_J={1/2}}\) and \(\ket{61P_{3/2},m_J={1/2}}\), the \(C_3\) coefficient becomes magnetic field dependent, because the expectation value of the electric dipole operator between the bare states \(\ket{61S_{1/2},m_J={1/2}}\) and \(\ket{61P_{1/2},m_J={1/2}}\) is different from the adiabatically connected eigenstates of the Hamiltonian in the magnetic field. This is the reason why in an magnetic field of \SI{185}{\gauss}, the \(C_3\) coefficient calculated using the pair interaction library is lowered to \(C_3=2\pi\times\SI{875}{\mega\hertz}\) (from \(C_3=2\pi\times\SI{1626}{\mega\hertz}\) at \(B=0\))  for the eigenstates adiabatically connected to the states \(\ket{61S_{1/2},m_J={1/2}}\) and \(\ket{61P_{1/2},m_J={1/2}}\), and increased to \(C_3=2\pi\times\SI{3876}{\mega\hertz}\) (from \(C_3=2\pi\times\SI{3151}{\mega\hertz}\) at \(B=0\)) for the eigenstates adiabatically connected to the states \(\ket{61S_{1/2},m_J={1/2}}\) and \(\ket{61P_{3/2},m_J={1/2}}\).
For two atoms in the same Rydberg state, exemplary for two atoms in the state \(\ket{61S_{1/2},m_J={1/2}}\), the energy has no corrections to perturbation theory up to the first order, as the electric dipole element between two atoms in the same level vanishes. As a consequence, the energy of this 2-level state changes in second order perturbation theory, i.e.
\begin{equation}
    \Delta E = \sum_{\alpha \beta} \frac{\mel{nS nS}{\hat{H}_{\mathrm{DDI}}}{\alpha \beta}\mel{\alpha \beta}{\hat{H}_{\mathrm{DDI}}}{nS nS}}{2E_{nS}-E_\alpha-E_\beta} = \frac{C_6}{r^6}
    \label{eq:VdW}
\end{equation}
In the absence of a magnetic field, all the different \(m_J\) states in an atomic \(\ket{nL_J}\) manifold are degenerate, and hence the sum over all different \(m_J\) states with the same \(n\), \(L\) and \(J\) quantum numbers for the two-pair states \(\ket{\alpha\beta}\) in the matrix element term is independent of direction. As a consequence, the change in energy is isotropic. Moreover, the dipole-dipole Hamiltonian has a \(r^{-3}\) scaling, hence the energy shift will scale as \(r^{-6}\). The proportionality constant is the \(C_6\) dispersion coefficient.

In a finite magnetic field, the energies \(E_\alpha\) and \(E_\beta\) are magnetic field dependent, and different \(m_J\) states are not degenerate anymore. This leads to an anisotropy in the energy shift, such that the energy shift
\begin{equation}
    \Delta E = \frac{C_6(\theta)}{r^6}
\end{equation}
is dependent on the angle between the interatomic axis with the magnetic field axis. Numerically evaluating equation (\ref{eq:VdW}) and using equation (\ref{eq:HDD}), we found convergence including states \(\ket{\alpha(\beta)}=\ket{nLJ_{mJ}}\) for \(n \in \{57,58,59,60,61,62,63,64,65\}\), \(L \in \{0,1,2,3\}\), all possible values of \(J\) for these states, and \(m_J \in \{-0.5,0.5,1.5\}\).
\begin{figure}[h]
    \centering
    \includegraphics[width=\linewidth]{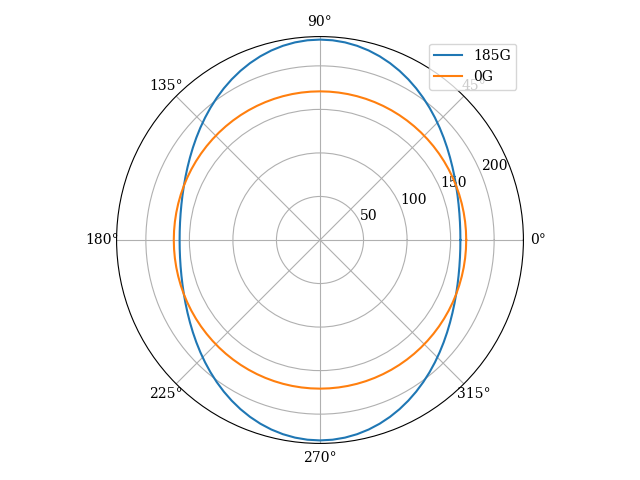}
    \caption{Polar plot of the \(C_6\) coefficient at different magnetic fields. Angle corresponds to angle between interatomic axis and magnetic field direction, distance to origin to the value of the \(C_6\) dispersion coefficient in \unit{\giga\hertz\micro\meter^6}. Orange curve: without magnetic field. Blue curve: At \SI{185}{\gauss}.
    }
    \label{fig:figure_appendix1}
\end{figure}
The results are shown in Fig. \ref{fig:figure_appendix1}. Unlike at zero magnetic field, we observe a strong increase of the interaction strength in the direction perpendicular to the magnetic field. This anisotropy in the Rydberg interaction introduces a positional anisotropy during excitation of atoms to the Rydberg state \(\ket{61S_{1/2},m_J=1/2}\). As a consequence, it is hard to determine the anisotropic Rydberg density in the experimental dipole trap. In order to estimate the median interaction strength \(J_{med}\) of the atomic cloud as described in \cite{signoles_glassy_2021}, we use the median value of \(C_6\) averaged over all directions, where we make a systematic error by assuming an isotropic Rydberg blockade.

\subsection{Time dependence of the ZFA and FA protocols}
\begin{figure}[!htb]
    \centering
    \includegraphics[width=\linewidth]{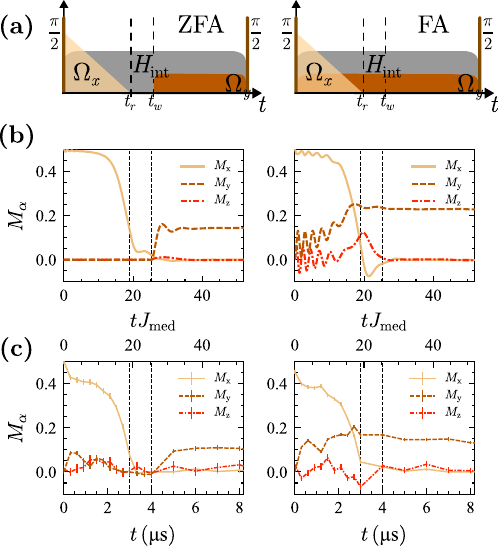}
    \caption{Time evolution of the different magnetization components \(M_{x}\) (light orange), \(M_{y}\) (brown) and \(M_{z}\) (red) as a function of time for the ZFA (left) and FA (right) protocols. Black lines indicate waiting time in the ZFA protocol. (a) Sketches of applied fields in different direction. Brown lines depict Rabi \(\pi/2-\)pulses.
    (b) Exact diagonalization simulation of the magnetization components. Evolution time \(t\) in units of inverse median interaction strength \(t J_\mathrm{med}\). (c) Experimental data
    ($\Omega_{\frac{\pi}{2}}/2\pi=\SI{12.18}{\mega\hertz},\,
    \Omega_x/2\pi=\SI{3.65}{\mega\hertz},\,
    \Omega_y/2\pi=\SI{0.49}{\mega\hertz},\,
    t_{r}=\SI{3}{\micro\second},\,
    t_{w}=\SI{1}{\micro\second}$)
    of magnetization components measured throughout the annealing protocol. Evolution time \(t\) both in \unit{\micro\second} (bottom axis) as well as in interaction cycles (top axis).
    }
    \label{fig:figure2}
\end{figure}

We prepare the system in the fully polarized state in \(z\)-direction by exciting the atoms to the Rydberg state \(\ket{\downarrow}\). A rectangular Rabi \(\pi/2-\)pulse of the microwave drive (Rabi frequency \(\Omega\)) realizes the fully polarized state in \(x-\)direction.
In the weak disordered regime, we observe all components of the magnetization as a function of time for both the ZFA and the FA protocol. The protocols are sketched in Fig. \ref{fig:figure2}a.
The 12 particle numerical simulation in Fig. \ref{fig:figure2}b is compared to the experimental results in Fig. \ref{fig:figure2}c.

We observe both in experiment and simulations that \(M_{x}\) and \(M_{z}\) have relaxed to zero at the end of the waiting time in the ZFC protocol. Since \(t_{w}\) is of the same order of magnitude as \(t J_\mathrm{med}\), we conclude that only a few interaction cycles after the ramp are sufficient for equilibration of the magnetization, in consistence with previous findings \cite{franz_absence_2022,franz_observation_2024}.
\SI{2}{\micro\second} after application of the probe field in the ZFA protocol (\SI{3}{\micro\second} after the end of the ramp in the FA protocol) \(M_{y}\) reaches an equilibrium value, which does not change over several interaction cycles and we call the ZFA (FA) magnetization.

Comparing the simulations and the experimental results, we observe two effects due to experimental imperfections:
In the ZFA protocol, oscillations in \(M_{y}\) and \(M_{z}\) occur during the annealing ramp in the experiment, which are absent in the simulation. As the amplitude of the microwave drive decreases, possible light shifts can create a detuning which results into an effective field in \(z-\)direction. Moreover, if the phase of the microwave drive deviates from \(\frac{\pi}{2}\), the annealing field does not point in \(x-\)direction, but has an additional \(y-\)component. Both effects create oscillations in the \(y-\) and \(z-\)component of the magnetization.
In the FA protocol, we note that \(M_{x}\) is less than 0.5 for the initial state, and has nonzero \(y-\) and \(z-\)components. We attribute this to fluctuations in the power of the microwave drive, which results into imperfect \(\frac{\pi}{2}\) pulses. Nonetheless, the prepared state has a large overlap with the ground state, such that the effect on the annealing scheme is small.
For the susceptibility measurements in the paper, we fix \(t_{w}=\SI{1}{\micro\second}\) and measure the ZFA/FA magnetization \SI{3}{\micro\second} after the annealing ramp.

\newpage
\bibliographystyle{apsrev4-2}
\bibliography{bib}